\documentclass[a4paper,twocolumn,11pt,accepted=2023-06-03]{quantumarticle}
\pdfoutput=1

\usepackage[utf8]{inputenc}
\usepackage[english]{babel}
\usepackage[numbers]{natbib}
\usepackage[T1]{fontenc}
\usepackage{amsmath}
\usepackage{graphicx}
\usepackage{tabularx}
\usepackage{booktabs}
\usepackage{color}
\usepackage[ruled,vlined]{algorithm2e}
\usepackage[thmmarks,amsmath,standard]{ntheorem}
\usepackage{physics}
\usepackage{float}
\usepackage{hyperref}
\usepackage{multirow}

\newcommand{\mc}[1]{\mathcal{#1}}
\newcommand{\genset}[1]{\langle#1\rangle}
\newcommand{\qwith}{\quad \mathrm{with} \quad}

\newcommand{\iden}{\vb{1}}
\newcommand{\defi}{:=}
\newcommand{\ifed}{=:}

\newcommand\ddfrac[2]{\frac{\displaystyle #1}{\displaystyle #2}}

\DeclareMathOperator*{\argmax}{arg\,max}

\begin{document}

\title{Generalized Belief Propagation Algorithms for Decoding of Surface Codes}

\author{Josias Old}
\email[]{j.old@fz-juelich.de}
\affiliation{Institute for Quantum Information, RWTH Aachen University, Aachen, Germany}
\affiliation{Institute for Theoretical Nanoelectronics (PGI-2), Forschungszentrum Jülich, Jülich, Germany}

\author{Manuel Rispler}
\email[]{rispler@physik.rwth-aachen.de}
\affiliation{Institute for Quantum Information, RWTH Aachen University, Aachen, Germany}
\affiliation{Institute for Theoretical Nanoelectronics (PGI-2), Forschungszentrum Jülich, Jülich, Germany}
\affiliation{QuTech, Delft University of Technology, Lorentzweg 1, 2628 CJ Delft, The Netherlands}
%

% \date{24/04/2023}

\begin{abstract}
    Belief propagation (BP) is well-known as a low complexity decoding algorithm with a strong performance for important classes of quantum error correcting codes, e.g. notably for the quantum low-density parity check (LDPC) code class of random expander codes. However, it is also well-known that the performance of BP breaks down when facing topological codes such as the surface code, where naive BP fails entirely to reach a below-threshold regime, i.e. the regime where error correction becomes useful. Previous works have shown, that this can be remedied by resorting to post-processing decoders outside the framework of BP. In this work, we present a generalized belief propagation method with an outer re-initialization loop that successfully decodes surface codes, i.e. opposed to naive BP it recovers the sub-threshold regime known from decoders tailored to the surface code and from statistical-mechanical mappings. We report a threshold of $17\%$ under independent bit-and phase-flip data noise (to be compared to the ideal threshold of $20.6\%$) and a threshold value of $14\%$ under depolarizing data noise (compared to the ideal threshold of $18.9\%$), which are on par with thresholds achieved by non-BP post-processing methods.
\end{abstract}
\maketitle

% \tableofcontents
%

\section{Introduction}
Quantum computing devices suffer from operational errors and decoherence. Methods to keep errors in check and advance towards fault-tolerant quantum computing involve \emph{quantum error correcting codes}. A code is formally defined as a \(k\)-dimensional subspace of some \(n\)-dimensional Hilbert space. 
In quantum error correcting codes called \emph{(quantum) stabilizer codes}, the error correction procedure consists of three main elements: the (non-destructive) measurement of stabilizer operators, the decoding of this measurement to obtain a suitable recovery operation, and the application of the latter to correct the errors on the codestate.
In order to achieve \emph{fault-tolerant quantum computation}, all three stages have to be done in a fast and efficient way to avoid the accumulation of errors. 

A class of codes which recently received a lot of attention are \emph{Quantum Low-Density Parity-Check codes} (qLDPC codes)~\cite{Breuckmann2021}. Given an asymptotically constant rate \(r = \frac{k}{n}\), they are proven to achieve fault-tolerance with only constant overhead. 
That is, the ratio of total number of qubits used in the fault-tolerant protocol to the number of qubits in a non-fault tolerant circuit is asymptotically constant for increasing code size~\cite{gottesman2014fault}.
An important quantity for error correction is the \emph{minimum distance}. It is the minimum weight (\textit{i.e.} the number of qubits involved) of a logical operation on the codespace. Naturally, a large distance implies a better protection against errors. 
Quantum error correcting codes are called \emph{good} if their rate is constant and the minimum distance scales linearly with increasing code size. The existence of such codes, however non-qLDPC, has been proven for a long time~\cite{calderbank1996good,ashikhmin2001asymptotically}. In a recent seminal work, Panteleev and Kalachev showed that it is actually possible to construct asymptotically good qLDPC codes~\cite{panteleev2022asymptotically}. Shortly after that breakthrough, several other constructions achieved a similar scaling behavior~\cite{leverrier2022quantum,lin2022good}.

A key ingredient of a fault-tolerant protocol, which will be the focus of this work, is an efficient decoding algorithm. Proofs of fault-tolerance often consider the classical processing of the error correction cycle as free. In practice, this is not the case and decoders should be fast enough to prevent additional errors from occurring. This typically results in a trade-off between decoder accuracy, i.e. how well the decoding algorithm finds a good, ideally the most likely correction and decoder computational complexity, i.e. how fast the algorithm can be executed as a function of the input size. There are a variety of decoders that are highly adapted to the quantum code in use.  
In this paper, we consider a generalization of a classical decoding algorithm called \emph{belief propagation} (BP) or \emph{sum-product algorithm}. For classical LDPC codes, it is among the best practical decoding algorithms: it allows to achieve error correction close the theoretical upper bound on the information transfer, the \emph{Shannon capacity}~\cite{mackay1997near}, while remarkably maintaining low algorithmic complexity. In principle, any classical decoding algorithm can be used to decode quantum codes, which stems from the fact that any quantum code can be formulated as a so-called CSS (Calderbank-Shor-Steane) code, which decouples into two classical codes that can (in principle) be decoded independently. However, this has many pitfalls. The most relevant one for us here is the observation that belief propagation on the surface code can completely fail to converge or converge to decoding decisions that do not remove the error. This is thought to be the underlying reason for the observation that the surface code produces no sub-threshold regime (the regime of error rate, where the logical error rate can be arbitrarily suppressed by increasing the code size) when decoding with BP~\cite{raveendran2021trapping}. This is in stark contrast to tailored quantum decoders such as minimum weight matching, under which the surface code is typically celebrated for its high threshold value~\cite{Raussendorf2007}. Recently introduced post-processing methods to BP showed that these can in principle decode various types of qLDPC codes~\cite{panteleev2019degenerate,roffe2020decoding}. We will focus on a single decoder that achieves the same at a lower complexity. The  Generalized Belief Propagation (GBP) algorithm was previously considered in the context of decoding quantum bicycle codes~\cite{raveendran2019syndrome}.

This paper is structured as follows. First, we review the basics of stabilizer error correction and quantum LDPC codes. In Section~\ref{sec:GBP} we show the relation of GBP to standard BP and adapt it for the decoding of quantum codes. Section~\ref{sec:GBP_surface} applies GBP to surface codes and shows numerical evidence of the emergence of a threshold.

\subsection{Stabilizer Codes}
Stabilizer codes are defined by an Abelian subgroup of the Pauli group, \(\mc{S} \subseteq \mc{P}_n\) with \(-\iden \notin \mc{S}\). The (commuting) elements of the group are the \emph{stabilizers} or \emph{parity checks} \(S \in \mc{S}\).
The codespace \(\mc{Q}\) is then defined as the subspace of the Hilbert space \(\mc{H}^{n}\) that is stabilized by \(\mc{S}\). For any codestate \(\ket{\psi}\) it therefore holds that \(S \ket{\psi} = \ket{\psi} \; \forall S \in \mc{S}\).
If there are \(n-k\) independent generators for \(\mc{S}\), \(\abs{\genset{\mc{S}}} = n-k\), then the codespace is  \(k\)-dimensional and encodes \(k\) qubits. 

Consider a Pauli error \(E \in \mc{P}_n\) occurring on a codestate, \(\ket{\psi} \to E \ket{\psi} \). Such an error can be detected by measuring all stabilizer generators, if any of those anticommute with the error, \(S E \ket{\psi} = - E S \ket{\psi}\) for some \(S \in \mc{S}\). The binary outcome of all stabilizer measurements is also called the \emph{syndrome} or \emph{syndrome vector} \(\vb{s} \in \mathrm{GF}(2)^{n-k}\) with
\begin{align}
        s_c = \langle E,S_c \rangle \defi \begin{cases}
                0 \qif \comm{E}{S_c} = 0, \\
                1 \qif \acomm{E}{S_c} = 0. \\
        \end{cases}
\end{align}
Here, \(s_c\) denotes the measurement outcome of stabilizer generator \(S_c\) for \(c = 1,\dots,n_c = n-k\).
This gives rise to three different scenarios. We say there occurred a
\begin{enumerate}
        \item \emph{detectable error} if \(\vb{s} \neq 0\),
        \item \emph{trivial error} if \(\vb{s} = 0\) and \(E \in \mc{S}\),
        \item \emph{logical error} if \(\vb{s} = 0\) and \(E \notin \mc{S}\).
\end{enumerate}
The last case represents the operators that map non-trivially between codestates, the \emph{logical operators}. They can formally be defined using the \emph{centralizer} of \(\mc{S}\) in \(\mc{P}_n\), \(\mc{C}_{\mc{P}}(\mc{S}) = \{L : LS = SL \; \forall S \in \mc{S}\}\) , such that the logical operators are \(\mc{L} = \mc{C}_{\mc{P}}(\mc{S}) \setminus \mc{S}\).
The \emph{(minimum) distance} of the stabilizer quantum code then corresponds to the minimal weight of a logical operator,
\begin{align}
        d = \min_{L \in \mc{L}} \abs{L}.
\end{align}

\subsection{Decoding of Stabilizer Codes}
The \emph{decoding problem} refers to the inference of a suitable correction from the measured syndrome. Because trivial errors have zero syndrome, they define an equivalence class for every detectable error. This feature called \emph{degeneracy} implies that corrections only need to be found up to a trivial error.
Due to the linearity of the codes, errors up to weight \(t = \lfloor \frac{d-1}{2} \rfloor\) can be uniquely matched to a codestate and hence be corrected for.
A simple decoder involves a lookup-table which stores a suitable correction, for example the lowest-weight error matching each measured syndrome. 
Making assumptions on the error probabilities can improve this approach. 

To that end, we consider Pauli error channels,
\begin{align}
    \mc{E}(\rho) = \sum_{E \in \mc{P}_n} p(E) E \rho E^\dagger.
\end{align}
Furthermore, we assume that the qubits are memoryless and suffer from errors independently,
\begin{align}
    p(E) = \prod_{q = 1}^{n_q} p(E_q).
\end{align}
We can write the probability of errors conditioned on the observation of a syndrome using Bayes' rule and the fixed "evidence" \(p(\vb{s}) = 1\)  as
\begin{align}
    p(E|\vb{s}) &= p(E) p(\vb{s}|E)  \\
           &= \prod_{q = 1}^{n_q} p(E_q) \prod_{c = 1}^{n_c} \delta(\langle E,S_c \rangle = s_c).
           \label{eqn:prob_error}
\end{align} 
By \(\delta(i=j)\) we denote the Kronecker delta \(\delta_{ij}\). It assigns zero probability to all error configurations \(E\) that have a commutation relation inconsistent with the measurements.

In quantum error correction, the ideal decoder returns an error guess, which is in the most likely error class, specified by all errors that are equivalent up to an element of the stabilizer group.  
Given a syndrome, this \emph{maximum likelihood decoding} identifies 
\begin{align}
    E^{\star} \in  \argmax_{E\mc{S}}  p(E\mc{S}| \vb{s}) = \argmax_{E \mc{S}} \sum_{S \in \mc{S}} p(ES| \vb{s}) .
\end{align}

Note, however, that directly calculating the probabilities for an error class (or even just storing the lookup-table) quickly becomes intractable since there are \(2^{(n-k)}\) different syndromes. For example, storing all syndromes of a code defined by \(42\) independent stabilizer generators requires approximately \(550 \mathrm{GB}\) of memory.

Possibly more efficient decoding strategies rely on relaxed constraints such as finding the  
\begin{itemize}    
    \item most likely error, \textit{i.e.} identifying 
    \begin{align}
        E^{\star} = \argmax_{E \in \mc{P}_n} p(E|\vb{s})
    \end{align}
\end{itemize}
or the
\begin{itemize}  
    \item qubit-wise most likely error, \textit{i.e.} identifying 
    \begin{align}
        E^{\star} = \{\argmax_{E_q \in \mc{P}_1} p_q(E_q|\vb{s})\}_{q=1}^{n_q},
    \end{align}
\end{itemize}
where \(p_q(E_q|\vb{s}) = \sum_{q' \neq q} p(E|\vb{s})\) are the single-qubit marginal probabilities.

Note that these equations are agnostic of the quantum nature of the underlying problem and have been studied extensively in various settings including classical decoding. Finding the most likely error is already less involved than finding the most likely error class, but is in general still NP-complete~\cite{berlekamp1978inherent}.
Calculating \(p(E|\vb{s})\) directly and even inferring the marginal probabilities \(p_q(E_q|\vb{s})\) still involves an exponential number of components. 
However, there exist algorithms that can - under certain conditions - calculate the marginals in linear complexity \(\mc{O}(n)\). This comes at the cost that the qubit-wise most likely error might globally be inconsistent with the observed syndrome.
Before introducing such an algorithm, the \emph{belief propagation} algorithm, we fix the notation and graphical representations used throughout this paper.

\subsection{Representation of Stabilizer Codes} \label{ssec:repr}
\paragraph*{Algebraic representation} The stabilizer group and most operations used in stabilizer error correction can be mapped from the Pauli group to vector spaces over finite fields (or \emph{Galois Fields}), denoted by \(\mathrm{GF}(q)\) with \(q = \{2,4\}\)~\cite{gottesman1997stabilizer}.

In both cases, the Pauli word \(E\) is mapped to a vector \(\vb{e} \in \mathrm{GF}(q)\) of length \((3 - \frac{q}{2})n\). The group operation is mapped to element-wise addition on the finite fields, \(E E' \mapsto \vb{e} + \vb{e}'\). Commutation of two Paulis \(E, E'\) can be checked using the \emph{symplectic product} denoted by \(\star\),
\begin{align}
        \langle E, E' \rangle \mapsto \vb{e} \star \vb{e}'.
\end{align}
The stabilizer generators are put in a \emph{parity check matrix} \(\vb{H} \in \mathrm{GF}(2)^{(n-k) \times 2n}\) or \(\vb{H} \in \mathrm{GF}(4)^{(n-k) \times n}\), such that the measurement of all stabilizers can be represented by the symplectic matrix-vector product
\begin{align}
        \vb{H} \star \vb{e} \ifed \vb{s}.
\end{align}
The actual implementation in the binary and quaternary framework can be found in App.~\ref{app:representation}.

\begin{figure}
\includegraphics[width=\linewidth]{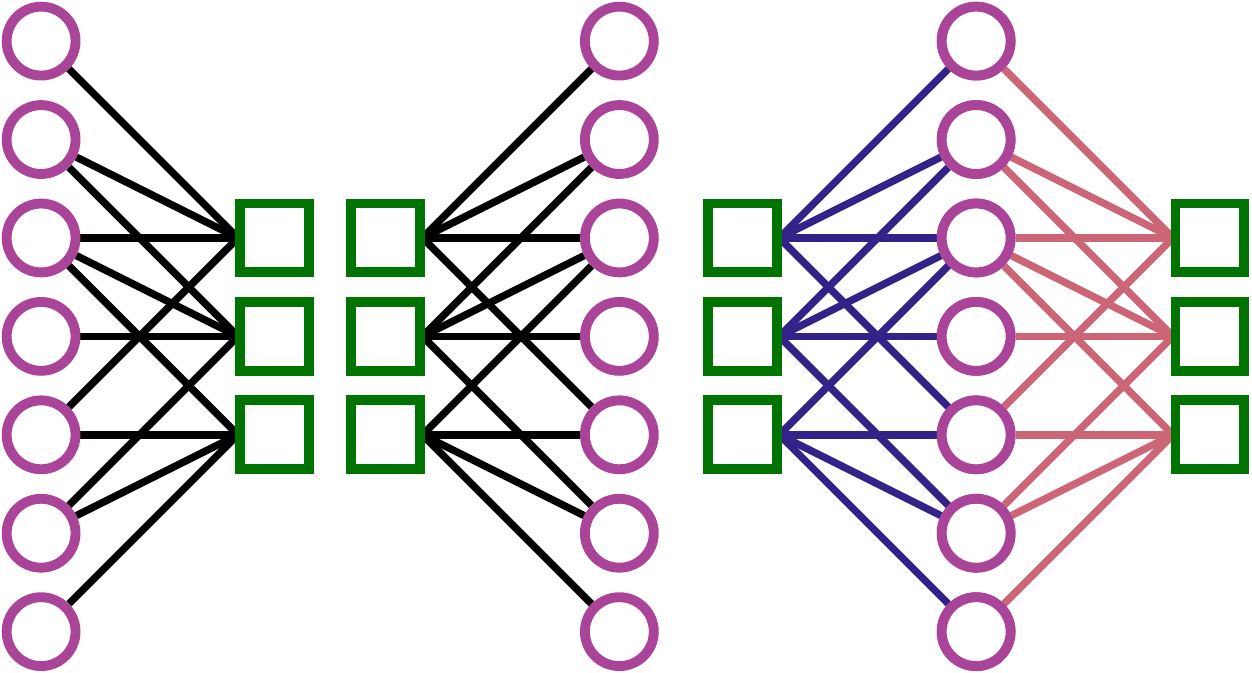}
\caption{\label{fig:tannergraph} Tanner graph representation of the Steane code in binary (left) and quaternary (right) representation. Purple circles correspond to qubits, green squares to parity checks. In the quaternary representation, blue edges correspond to \(X\)-type checks and red edges to \(Z\)-type checks. Note that since the Steane code is a CSS-code, the binary Tanner graph splits into two disjoint graphs \(\mc{T} = \mc{T}_X \sqcup \mc{T}_Z\).}
\end{figure}

\paragraph*{Tanner graph representation}
Stabilizer codes can be graphically represented as \emph{Tanner graphs}, similar to classical codes~\cite{tanner1981tanner}. These are bipartite graphs with two vertex sets \(Q\) and \(C\) representing the qubits and the stabilizer measurements/ parity checks respectively. The edge set \(E\) consists of edges \(e = (q,c)\) drawn between vertices \(q \in Q\) and \(c \in C\) if qubit \(q\) is involved in the parity-measurement of stabilizer \(c\). Different types of Paulis can be distinguished by coloring the edges. With this correspondence, the parity-check matrix \(\vb{H}\) is the \emph{biadjacency matrix} or \emph{reduced adjacency matrix} of the Tanner graph \(\mc{T} = (Q \cup C,E)\).

\paragraph*{Example: The Steane Code}
The Steane code~\cite{steane1996multiple} is a \([[7,1,3]]\)- quantum code with stabilizer generators
    \begin{align}
        \genset{\mc{S}} = \{&X_0 X_1 X_2 X_4, X_1 X_2 X_3 X_5, X_2 X_4 X_5 X_6, \\
        &Z_0 Z_1 Z_2 Z_4, Z_1 Z_2 Z_3 Z_5, Z_2 Z_4 Z_5 Z_6\}.
    \end{align} 

In the algebraic representations, the parity check matrices are
\begin{align}
\vb{H}_{\mathrm{GF}(2)} = \mqty[\vb{H}_X & 0 \\ 0 & \vb{H}_Z] = \smqty[\dmat{\smqty{1&1&1&0&1&0&0\\
0&1&1&1&0&1&0\\
0&0&1&0&1&1&1} ,\smqty{1&1&1&0&1&0&0\\
0&1&1&1&0&1&0\\
0&0&1&0&1&1&1}}] 
\end{align}
and 
\begin{align}
        \vb{H}_{\mathrm{GF}(4)} = \mqty[\vb{H}_X \\ \omega \vb{H}_Z] = \smqty[\smqty{1&1&1&0&1&0&0\\
        0&1&1&1&0&1&0\\
        0&0&1&0&1&1&1\\ 
        \omega&\omega&\omega&0&\omega&0&0\\
        0&\omega&\omega&\omega&0&\omega&0\\
        0&0&\omega&0&\omega&\omega&\omega}]
\end{align}
and Tanner graphs are shown in Fig.~\ref{fig:tannergraph}.

\subsection{Low-Density Parity-Check Codes}
Low-Density Parity-Check (LDPC) codes are families of classical codes with a sparse parity-check matrix. %
The most successful classical LDPC codes rely on random or pseudo-random constructions of the parity-check matrix, most notably Sipser and Spielman's \emph{Expander Codes}~\cite{sipserspielman1996expander}. Their properties include a constant rate, a linear distance and efficient decoders, which is often referred to as \emph{good code}~\cite{mackay2003information}. 
This has led to try and construct quantum versions of LDPC codes (qLDPC codes) in the hope of obtaining \emph{good} quantum codes with a constant rate and distance linear in the number of qubits. In addition to having good decoding properties, it was famously shown by Gottesman that such codes, if they exist, enable fault-tolerant quantum computation with only constant overhead~\cite{gottesman2014fault}. 

In general, qLDPC codes can be defined similarly to classical LDPC codes as codes with a sparse parity check matrix. This corresponds to quantum stabilizer codes with stabilizer generators of low weight that is upper bounded by a constant. In particular, a \((d_c,d_q)\)-qLDPC code ensemble has parity checks measuring at most \(d_c\) qubits and every qubit is involved in at most \(d_q\) syndrome measurements. 
This broad definition includes a range of well known codes like the surface codes. They are defined on a lattice, therefore exhibit a high degree of symmetry and have only nearest-neighbor interaction~\cite{kitaev1997surface}. They have a minimum distance \(d \propto \sqrt{n}\) but suffer from a vanishing rate \(r \to 0\) as the number of qubits \(n \to \infty\).
A more general construction, the hypergraph product (HGP) codes, can achieve a constant rate \(r \to 1 - \frac{d_c}{d_q}\), when based on good (e.g. random) classical codes~\cite{tillich2009qldpc}. 

Very recently, a construction that builds on a \emph{\(G\)-lifted product} of expander codes over non-abelian groups \(G\) by Panteleev and Kalachev were proven to achieve constant rate and linear distance~\cite{panteleev2022asymptotically}. Similar constructions like the \emph{Quantum Tanner Codes} or codes from balanced product of lossless expanders achieve the same~\cite{leverrier2022quantum,lin2022good}. %

The advantageous properties of good qLDPC codes manifest at large qubit numbers. For near future applications, a moderate number of qubits in the order of a few hundred is realistic. It is therefore reasonable to focus on the less intricate hypergraph product construction, which we will briefly recap in the following. 

\paragraph*{Hypergraph Product Codes} \label{sec:ldpc_hgp}
The hypergraph product construction uses graph based arguments to derive quantum codes from classical codes. For details refer to~\cite{tillich2009qldpc}, in the following considerations the construction rule for the parity check matrices shall be sufficient.

Let \mbox{\(\vb{H}_1 \in \mathrm{GF}(2)^{m_1 \times n_1}, \vb{H}_2 \in \mathrm{GF}(2)^{m_2 \times n_2}\)} be parity check matrices of classical codes \(\mc{C}_1, \mc{C}_2\) with dimension \(k_1\) and \(k_2\). The Hypergraph Product (quantum) Code \(\mc{Q} = \mathrm{HGP}(\mc{C}_1,\mc{C}_2)\) with parity check matrix \(\vb{H}\) is a quantum CSS code with parameters
\begin{align}
        \qty[[n_1 n_2 + (n_1 - k_1)(n_2 - k_2), k_1 k_2, \min(d_1,d_2)]]
\end{align}
which has its parity check matrix constructed from the classical parity check matrices as
\begin{align}
        \vb{H} &= \mqty(\vb{H}_X & 0 \\ 0 & \vb{H}_Z), \\
        \vb{H}_X &= \mqty(\iden_{n_1} \otimes \vb{H}_2 & \vb{H}_1^T \otimes \iden_{m_2}), \\
        \vb{H}_Z &= \mqty(\vb{H}_1 \otimes \iden_{n_2} & \iden_{m_1} \otimes \vb{H}_2^T).
        \label{eqn:defihgp}
\end{align}
If we choose the base codes to have minimum distance linear in its length, the HGP construction gives quantum codes with minimum distance \(\Omega(\sqrt{n})\). The construction trivially preserves sparsity and therefore translates a classical LDPC property to a quantum LDPC property. 
Some choices of base codes give well known quantum codes.

\begin{figure}
    \includegraphics[width=\linewidth]{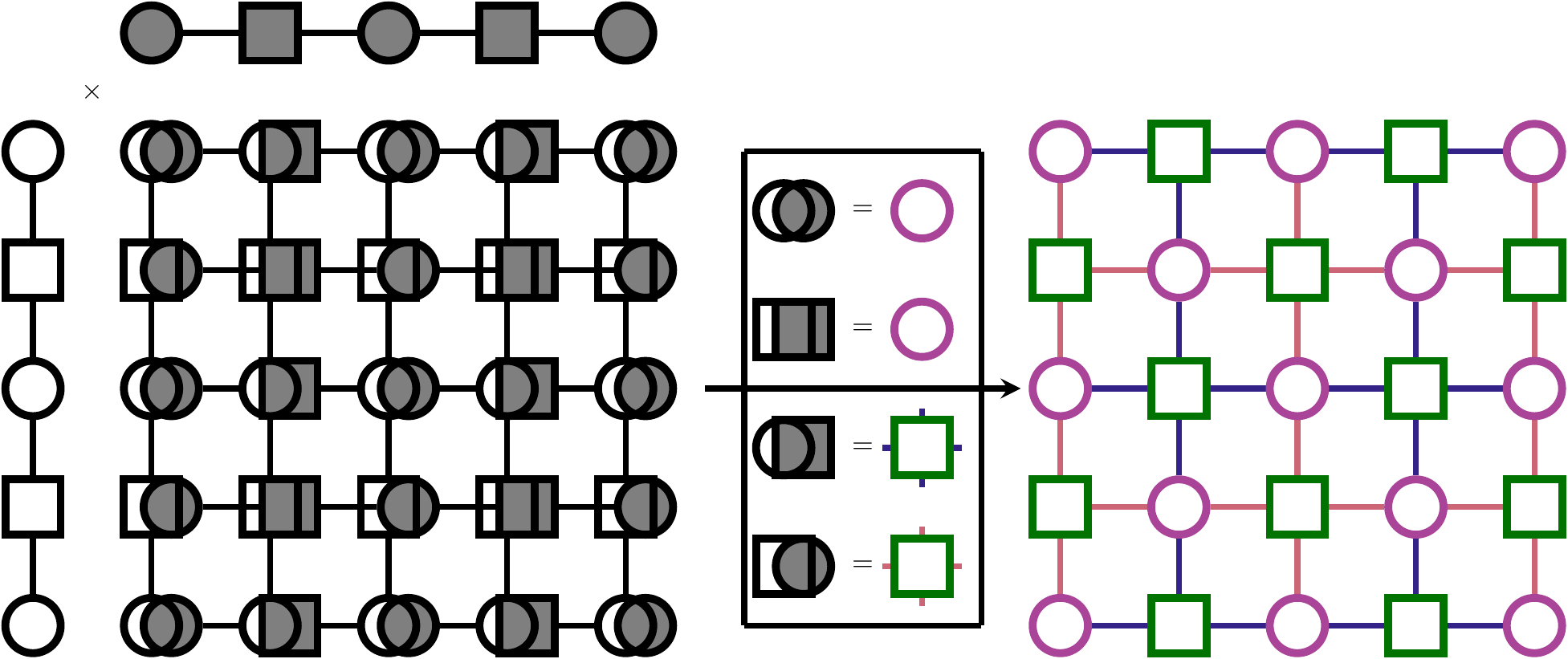}
    \caption{\label{fig:rep2surface} Graphical representation of the hypergraph product construction.The Cartesian graph product of the Tanner graphs of two repetition codes (left) yields the distance-\(3\) surface code (right) using the rules shown in the middle and explained in app.~\ref{app:hgp}}
\end{figure}

\begin{itemize}
    \item Taking the classical (cyclic) \emph{repetition codes} as base gives the topological \emph{(toric) surface codes}~\cite{kitaev1997surface}. They have vanishing rate and a distance \(d \propto \sqrt{n}\). The graph based construction is shown in Fig.~\ref{fig:rep2surface}.
    \item Taking the product of two (good) \emph{classical expander codes} yields the \emph{quantum expander codes}~\cite{leverrier2015quantumexpander}. These codes have constant rate and a minimum distance \(d \propto \sqrt{n}\). A slightly simplified version uses random classical codes that with, some known probability, have specific expansion properties. For such codes, a well known and widely used method for decoding is \emph{belief propagation}.
\end{itemize}

\section{Generalized Belief Propagation}\label{sec:GBP}
We now introduce Generalized Belief Propagation due to Yedidia, Freeman and Weiss (YFM)~\cite{yedidia2000generalized}. We then show how a decoder for quantum codes can be constructed from that.

The Tanner graph introduced in Sec.~\ref{ssec:repr} can also be thought of as an instance of a \emph{factor graph} representing a joint probability distribution over \emph{factors} \(f\)~\cite{frey1997factor},
\begin{align}
    p(\vb{x}) = \frac{1}{Z} \prod_{c = 1}^{n_c} f_c(\vb{x}_c).
\end{align}
In a physical system in thermal equilibrium,  \emph{Boltzmann's Law} gives the probability of a state, 
\begin{align}
    p(\vb{x}) = \frac{1}{Z} e^{-\beta E(\vb{x})} \qwith Z = \sum_{\vb{x} \in P} e^{-\beta E(\vb{x})}.
    \label{eqn:boltzmann}
\end{align}
\(P\) is the space of all possible states \(\vb{x}\) and \(\beta\) the inverse temperature which we set to \(1\) in the following.
A connection between the two can be drawn by identifying the probability distributions and therefore defining an energy \(E(\vb{x})\) of a state \(\vb{x}\) of the factor graph to be
\begin{align}
    E(\vb{x}) = - \sum_{c = 1}^{n_c} \ln{f_c}(\vb{x}_c).
\end{align}

The ultimate goal in quantum decoding is to maximize the probability distribution of the errors, which can be seen to correspond to finding the minimum free energy configuration. While this generally will be computationally unfeasible, one can construct a tractable variational ansatz. In the following, we will present GBP appealing to a general intuition for variational methods and refer the reader to the appendix~\ref{app:Variational} for background on variational methods and a detailed derivation of GBP. %To that end, we will first motivate the method of (G)BP by giving a review of variational methods. 

\subsection{Derivation of the GBP Algorithm}
Generalized Belief Propagation relies on region-based approximations to the free energy. These are a class of approximations to \(F(b)\), where the approximate free energy is a function of beliefs over sets of variables, called \emph{regions}. 
From now on, we will call the factor graph the \emph{Tanner graph}, the factors \emph{check nodes} and variables \emph{qubit nodes}, to facilitate the transition to the decoder based on GBP.

We can define a region \(r\) of a Tanner graph as a set of qubit nodes \(\mc{Q}_r\) and a set of check nodes \(\mc{C}_r\) such that if a check node \(c\) is in \(\mc{C}_r\), then all neighboring qubit nodes \(\{\Gamma(c)\}\) are in \(\mc{Q}_r\). 
The general idea is to define regions of the factor graph and then approximate the overall free energy with the sum of the free energies of all the regions, subject to the conditions ensuring validity which are shown in the following.

The thermodynamic quantities of a region are defined as
\begin{itemize}
    \item region energy
    \begin{align}
        E_r (\vb{x}_r) = - \sum_{c \in \mc{C}_r} \ln[f_c(\vb{x}_c)] - \sum_{q \in \mc{Q}_r} \ln[p(x_q)] 
    \end{align}
    \item region average energy and region entropy
    \begin{align}
        U_r (b_r) &= \sum_{\vb{x}_r} b_r(\vb{x}_r) E_r(\vb{x}_r) \\
        S_r (b_r) &= - \sum_{\vb{x}_r} b_r(\vb{x}_r) \ln[b_r(\vb{x}_r)]
    \end{align}
    \item region free energy
    \begin{align}
        F_r (b_r) = U_r (b_r) - S_r (b_r).
    \end{align}
\end{itemize}

When constructing regions, every check and qubit node should be contained in some region. Since check and qubit nodes might appear in multiple regions, it is necessary to introduce \emph{counting numbers}  \(c_r\) in order to ensure that every check and qubit is only counted once when summing over regions. They need to be chosen such that for a set of regions \(\mc{R}\) of a Tanner graph
\begin{align}
    \sum_{r \in \mc{R}} c_r \; \delta(c \in \mc{C}_r) = 1 \: \: \forall c \qc \sum_{r \in \mc{R}} c_r \; \delta(q \in \mc{Q}_r) = 1 \: \: \forall q.
    \label{eqn:counting_number_cond}
\end{align}
The overall, region-based approximate thermodynamic quantities are then given by
\begin{itemize}
    \item region-based average energy and region-based entropy \begin{align}
        U_{\mc{R}} (\{b_r\}) &= \sum_{r \in \mc{R}} c_r \; U_r(\{b_r\}), \\
        S_{\mc{R}} (\{b_r\}) &= \sum_{r \in \mc{R}} c_r \; S_r(\{b_r\}),
    \end{align}
    \item region-based free energy
    \begin{align}
        F_{\mc{R}} (\{b_r\}) = U_{\mc{R}} (\{b_r\}) - S_{\mc{R}} (\{b_r\}).
    \end{align}
\end{itemize}
Note that not for every choice of regions, a valid set of counting numbers can be found. To understand that, consider a valid choice of regions \(\mc{r} = \{r_i\}\) with the corresponding counting numbers \(\{c_i\}\), such that for qubit \(q\), Eq.~\ref{eqn:counting_number_cond} holds. Now adding any qubit \(q\) to a region \(r'\) with counting number \(c_{r'}\) and \(q \notin r'\) results in 
\begin{align}
    \sum_{r \in \mc{R}} c_r \; \delta(q \in \mc{Q}_r) = 1 + c_{r'},
\end{align}
which is only true iff \(c_{r'} = 0\).
Also note that different choices of regions can yield approximations of different quality~\cite{welling2012choice}. 

\subsubsection{Region graph}
\begin{figure}
    \includegraphics[width=\linewidth]{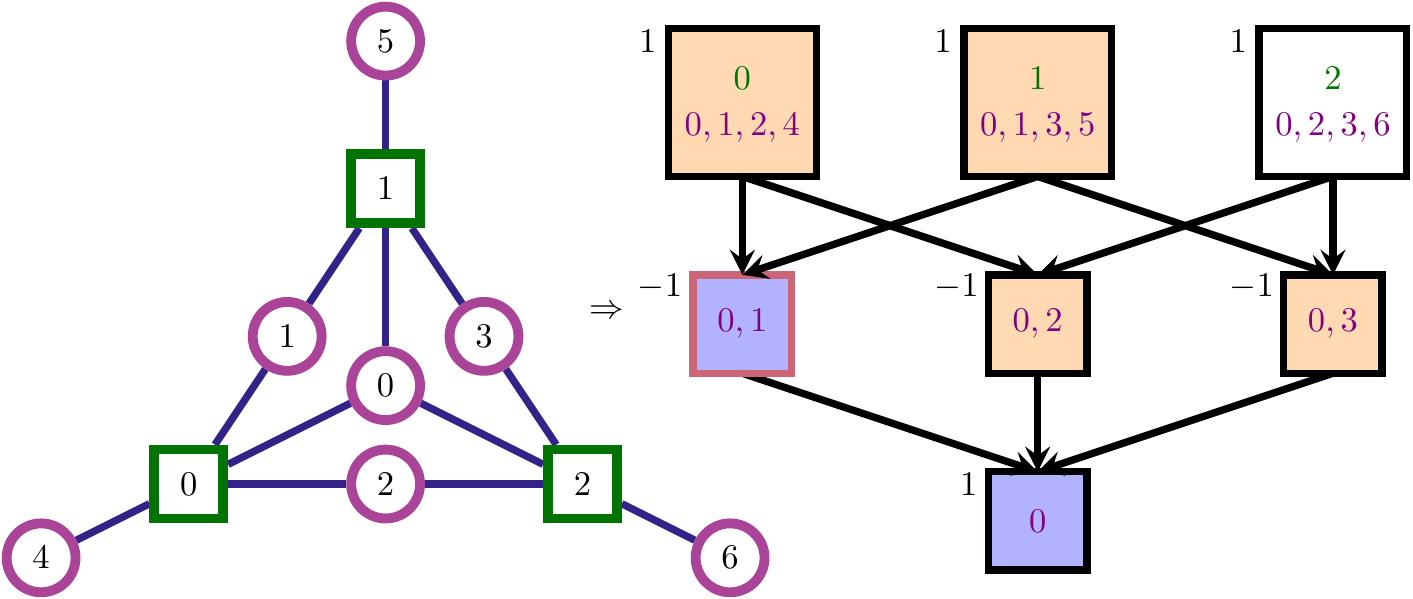}
    \caption{Example of a region graph for one part of the Steane code (corresponding to the classical Hamming code). Green and violet numbers represent checks and qubits respectively, counting numbers of the regions at top left corner. The "shadow" of the red bordered region containing qubits \(0,1\) is shaded in blue, the "blanket" in orange. See~\ref{sss:Notation} for the definitions of shadow and blanket. \label{fig:steane_rg}}
\end{figure}

Similarly to the Tanner graph, the relations of different regions can be formalized in a graph theoretical framework. To that end, YFM introduce the \emph{region graph} (RG). It is a labeled, directed graph \(\mc{G} = (\mc{R},\mc{E},L)\) in which each vertex corresponds to a region \(r \in \mc{R}\). 
A directed edge \(e \in \mc{E}\) may exist from region \(r_p\) to \(r_c\) if \((\mc{Q}(r_c) \cup \mc{C}(r_c)) \subset (\mc{Q}(r_p) \cup \mc{C}(r_p)) \), \textit{i.e.} if the set of constituents of the \emph{child} is a subset of the \emph{parent}s constituents. An example of a valid region graph for the Steane code is shown in Fig.~\ref{fig:steane_rg}.

\subsubsection{Notation} \label{sss:Notation}
We adopt the notation from~\cite{fanaskov2020gaussianBP} (which differs slightly from YFM~\cite{yedidia2000generalized}) and adapt it to the error correction setting,
\begin{itemize}
    \item \(\mc{R}\): set of all vertices (=regions) of the region graph,
    \item \(\mc{E}\): set of all (directed) edges of the region graph,
    \item \(P(r), C(r), A(r), D(r)\): set of all parents, children, ancestors, descendants of region \(r\),
    \item \(\mc{Q}_r,\mc{C}_r\): qubits and checks in region \(r\),
    \item \(S(r) = D(r) \cup r\): "shadow" of the region \(r\),
    \item \(B(r) = P\left(S(r)\right) \setminus S(r)\): "blanket" of the region \(r\).
\end{itemize}

\subsubsection{Algorithm} \label{sss:algorithm}
We can recover the beliefs as approximations of the true marginal probabilities by minimizing the region-based free energy \(F_{\mc{R}} (\{b_r\})\). To that end, a Lagrangian is constructed with the constraint that the beliefs shall be consistent between every parent and child region, \textit{i.e.}
\begin{align}
    \forall r, p \in \mc{R}, r \subset p \implies \sum_{\vb{x}_p \setminus \vb{x}_r} \vb{b}_p (\vb{x}_p) = \vb{b}_r (\vb{x}_r), 
    \label{eqn:consistency}
\end{align}
and a normalization constraint
\begin{align}
\forall r \in \mc{R}\qc \sum_{\vb{x}_r} \vb{b}_r (\vb{x}_r) = 1.
\end{align}
Setting the derivatives of the Lagrangian with respect to the beliefs equal to zero gives implicit equations for the Lagrange multipliers and the beliefs, as shown in~\cite{yedidia2005constructing}. They can be solved iteratively and for that reason, the Lagrange multipliers (or functions thereof) are often called \emph{messages} and the corresponding algorithms are referred to as \emph{message-passing algorithms}~\cite{mackay2003information}.

In the following we show a more intuitive approach. For that, assume that a belief of a region first contains all \emph{local} factors in that region. Messages from parent regions \(p\) into child regions \(r\) will be of the form \(m_{p \to r}(\vb{x}_r)\). In order to catch all possible dependencies, we consider all messages that include some variables that are contained in that region. However, over-counting of factors should be prevented.
This can be achieved by first considering all messages from parents into \(r\). Secondly, we take into account all messages from regions which are not descendants of \(r\) but point into its descendants \(D(r)\). This corresponds to the shadow and blanket of a region, such that an ansatz for the belief of a region \(r\) can be written as
\begin{align}
\vb{b}_r(\vb{x}_r) \propto \prod_{q \in \mc{Q}_r} p_q(x_q)  \; \prod_{c \in \mathcal{C}_r} f_c(\vb{x}_c) \prod_{\substack{a \in B(r),\\ b \in S(r)}} m_{a \to b}(\vb{x}_b).
\label{eqn:region_belief_general}
\end{align} 
By \(\vb{x}_{c}\) we denote the variables in the support of check \(c\).
The message update rules follows from demanding consistency between parent and child regions (Eq.~\ref{eqn:consistency}), giving the \emph{Parent-to-Child-Algorithm}. In the following, the upper index \((i)\) denotes the iteration step in order to formalize the iterative procedure. Note that with a uniform initialization of the messages, the beliefs of parent- and child regions \(p,r\) are incompatible at step \(i = 0\). We therefore update the message from \(p\) to \(r\) by that mismatch, 

\begin{align}
        \ddfrac{  \sum_{\vb{x}_p \setminus \vb{x}_r} \vb{b}^{(i)}_p (\vb{x}_p) }{\vb{b}^{(i)}_r (\vb{x}_r)} &\xrightarrow{i \to \infty} 1 \\
        \implies  m_{p \to r}^{(i+1)}(\vb{x}_r) &= m_{p \to r}^{(i)}(\vb{x}_r) \ddfrac{  \sum_{\vb{x}_p \setminus \vb{x}_r} \vb{b}^{(i)}_p (\vb{x}_p) }{\vb{b}^{(i)}_r (\vb{x}_r)}.
        \label{eqn:gbp_message_update}
\end{align}
The overlap of messages in the numerator and denominator can then be canceled out to reduce the number of calculations. 
In that form, it becomes clear that consistent beliefs correspond to converged messages.

\subsection{GBP as a Decoder for Quantum Codes}
In order to draw the connection to quantum decoding, we identify the variables with the qubits. The factors, \textit{i.e.} functional relations between qubits correspond to Kronecker delta tensors with entries according to the measured syndrome outcomes,
\begin{align}
    f_c(\vb{x}_c) = \delta[\vb{H}_c \star \vb{x}_{c} = s_c].
\end{align}
The state of the system \(\vb{x}\) coincides with the Pauli error \(E\) such that in the binary representation  \(\vb{x} = (x_1,x_2,\dots,x_{2 n_q}) \in \mathrm{GF}(2)^{2n}\) with \(x_q \in \{0,1\}\). In the following, we denote the errors by \(\vb{x}\) in order to avoid confusion with the energy.

\begin{figure}
    \centering
    \includegraphics[width=\linewidth]{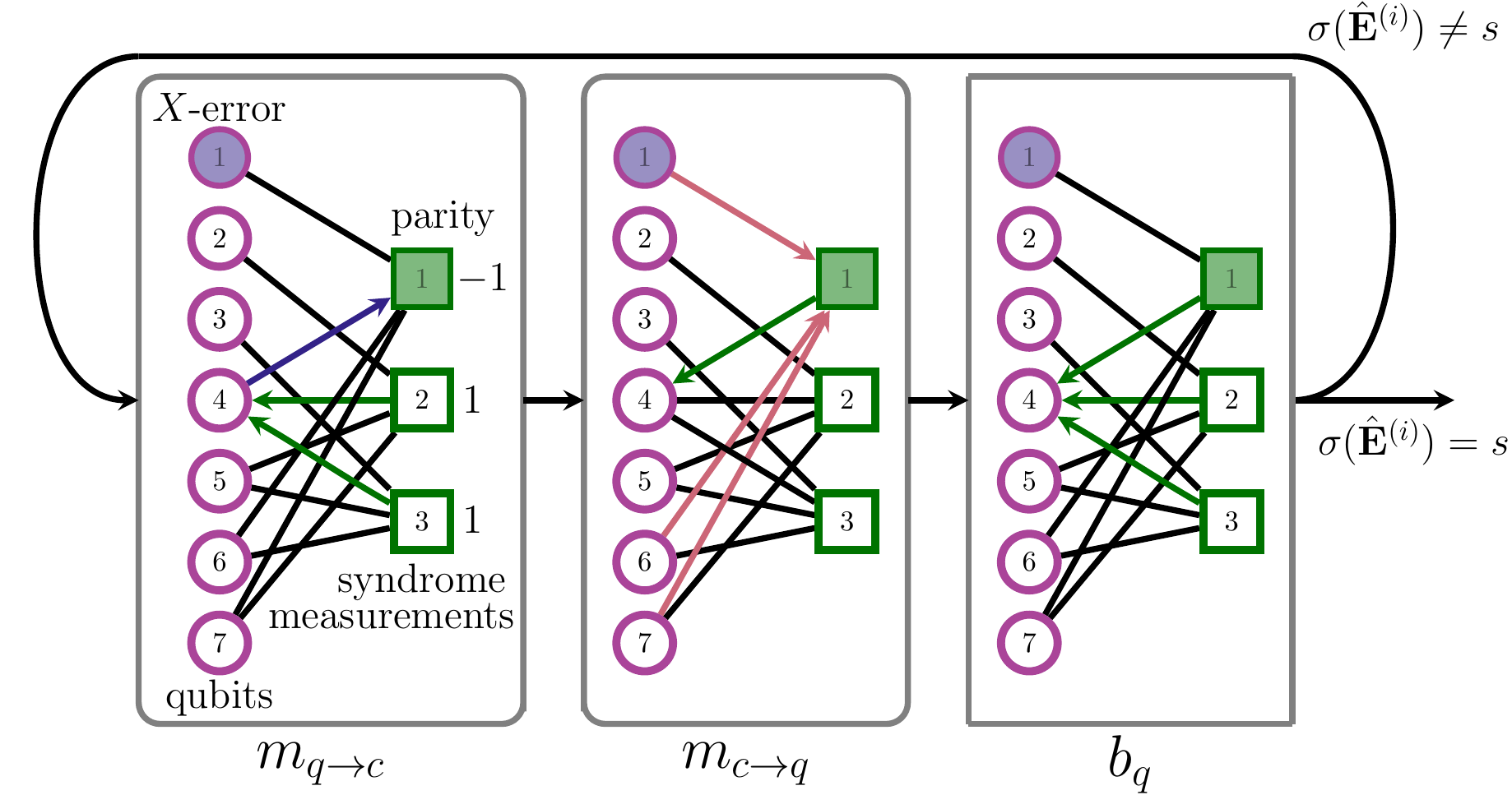}
    \caption{Belief Propagation decoding procedure. Shown is the \(X\)-Tanner graph of the \([[7,1,3]]\)-Steane code. An error on the first qubit violates parity check 1. Messages are indicated as arrows.
    From left to right: qubit to check messages based on incoming messages, check to qubit messages, marginal probabilities or belief and hard decision.}
    \label{fig:bp_messages}
\end{figure}

Note that the algorithm introduced above includes both an implementation separating the \(X\)- and \(Z\)- part of a CSS quantum code (\(\mathrm{GF}(2)\)) and a combined (\(\mathrm{GF}(4)\))- implementation. The differences then lie in the length of the vectors, the set of messages used to calculate region beliefs and in the syndrome function. The latter uses implementations of the symplectic product Eq.~\ref{eqn:sym_gf2} and Eq.~\ref{eqn:sym_gf4} in the \(\mathrm{GF}(2)\)- and \(\mathrm{GF}(4)\)- framework respectively.

At each iteration, the parent-to-child algorithm defined by Eq.~\ref{eqn:gbp_message_update} gives an estimate of the marginal probability distribution of region \(r\). These beliefs can be used to infer a guess of the error inflicted on the qubits. 

\subsubsection{How to make a hard decision}
The argumentum maximum (argmax) of the belief of a region gives the most probable error on the region's qubits. This is often referred to as \emph{making a hard decision}.
In the end, a hard decision has to be taken for every qubit individually, whereas every qubit might be part of multiple regions. 
Consistency of hard decisions from different regions is only guaranteed if all messages are converged. This is in general not the case while iterating and also not guaranteed to happen at all.
We therefore have to find a strategy to combine the different contributions from the region beliefs to the overall hard decision. A straightforward strategy to choose a hard decision is to focus on the highest level regions \(\mc{R}_0 : \{r \in \mc{R}: c_r = 1\}\) and make a hard decision as 
\begin{align}
    \hat{\vb{x}}^{(i)} = \bigcup_{r \in \mc{R}_0} \argmax_{\vb{x}_r} \vb{b}_r^{(i)}(\vb{x}_r).
    \label{eqn:gbp_hard_decision}
\end{align}
Naturally if the region-beliefs are not compatible, the overall error guess will not be consistent with the observed syndrome. 
In order to improve upon that, we find a more promising strategy.
Whenever the error guesses from different regions on a single qubit are incompatible, we compare the beliefs of their respective regions and settle for the one with the largest belief. 
This corresponds to decoding as
\begin{align}
    \hat{\vb{x}}^{(i)} &= \{\hat{\vb{x}}^{(i)}_q\}_{q=1}^{n_q} , \nonumber \\
    \hat{\vb{x}}^{(i)}_q &= \max_{x_q \in \vb{b}_r^{(i)}(\vb{x}_r)} \argmax_{\vb{x}_r} \vb{b}_r^{(i)}(\vb{x}_r).
    \label{eqn:gbp_hard_decision_2}
\end{align}

In the following, we first show how standard Belief Propagation can be recovered from this more general approach and then show two different strategies to cope with further obstacles.

\subsubsection{Bethe approximation} \label{sss:bethe_approximation}
One choice of regions called \emph{Bethe approximation} gives an approximation equivalent to the standard BP algorithm. 
It was originally introduced as \emph{sum-product decoding} for (classical) LDPC codes by Gallager~\cite{gallager1962ldpc}. Later it was independently rediscovered by Pearl as a method to efficiently calculate single variable marginals on factor trees~\cite{pearl1982reverend}. 
Poulin and Chung first applied BP to the decoding of quantum codes~\cite{poulin2008iterative}. For an introduction on the BP algorithm, see for example~\cite{rigby2019modifiedbp,roffe2020decoding}. 
Here, we show how to get the standard BP equations from the generalized ansatz.

For this purpose, construct two types of regions, large and small. The large regions each contain a single check node and its neighboring qubit nodes. The small regions comprise single qubit nodes such that all qubit nodes that have more than one parent have their own small region. The counting numbers are \(c_{r,\text{large}} = 1\) and \(c_{r,\text{small}} = 1 - \sum_{q \in \mc{Q}_r}\abs{P(q)}\). 
The beliefs of small regions \(\{l_i\}\) and large regions \(\{L_i\}\) are given by
\begin{align}
    \vb{b}_{c \in \{L_i\}}(\vb{x}_c) &{\propto}\hspace{-0.5em}  \prod_{q \in \mc{Q}_c} p_q(x_q) \delta[\vb{H}_c{\star}\vb{x}_{c}{=}s_c]\hspace{-1em}  \prod_{\substack{a \in P[C(c)]\setminus c \\ b \in C(c)}}\hspace{-1em} m_{a \to b}(x_b) \\
    \vb{b}_{q \in \{l_i\}}(x_q) &\propto p_q(x_q) \prod_{c \in P(q)} m_{c \to q}(x_q). \label{eqn:gpb_bethe_small_belief}
\end{align}
Using the consistency constraint Eq.~\ref{eqn:gbp_message_update} we find for the message updates
\begin{widetext}
\begin{align}
    m_{c \to q}^{(i+1)}(x_q) &= m_{c \to q}^{(i)}(x_q) \ddfrac{\sum_{\vb{x}_c \setminus x_q } \prod_{q' \in \mc{Q}_c} p_{q'}(x_{q'}) \delta[\sigma(\vb{x}_{\mc{Q}_c}) = s_c] \prod_{a \in P[C(c)]\setminus c, b \in C(c)} m_{a \to b}^{(i)}(x_b) }{p_q(x_q) \prod_{a \in P(q)} m_{a \to q}^{(i)}(x_q)} \\
    &= \ddfrac{\sum_{\vb{x}_c \setminus x_q } \prod_{q' \in \mc{Q}_c \setminus q} p_{q'}(x_{q'}) \delta[\sigma(\vb{x}_{\mc{Q}_c}) = s_c] \prod_{a \in P[C(c)]\setminus c, b \in C(c)} m_{a \to b}^{(i)}(x_b) }{\prod_{a \in P(q) \setminus c} m_{a \to q}^{(i)}(x_q)}\\
    &=  \sum_{\vb{x}_c \setminus x_q } \prod_{q' \in \mc{Q}_c \setminus q}  \delta[\sigma(\vb{x}_{\mc{Q}_c}) = s_c] \underbrace{p_{q'}(x_{q'}) \prod_{c' \in P(q') \setminus c} m_{c' \to q'}^{(i)}(x_{q'})}_{\ifed m_{q \to c}^{(i)}(x_q)}. \label{eqn:message_update_reduced}
\end{align}
\end{widetext}
The definition of the second type of messages (qubit to check) in the last line recovers the initial BP equations~\ref{eqn:bpq2c},~\ref{eqn:bpc2q} when identifying \(\mc{Q}_c \equiv \Gamma(c)\) and \(P(q) \equiv \Gamma(q)\).
Note that, instead of following rule Eq.~\ref{eqn:gbp_hard_decision} or~\ref{eqn:gbp_hard_decision_2}, standard BP thresholds the beliefs of the small regions Eq.~\ref{eqn:gpb_bethe_small_belief}.

Intuitively, the BP algorithm operates on the Tanner graph of the code by sending messages along its edges. Starting on the qubit site, it assigns initial probabilities of error (e.g. depending on the error channel) to the qubit nodes. This information is sent to the parity check nodes along the edges. 
The parity check nodes collect all incoming messages and send back a message to all adjacent qubits. In its components, this message contains the sum over all configurations compatible with the observed syndrome, excluding the target qubit.
Subsequently, the marginal probability of the qubits is calculated according to the incoming messages. If not compatible or converged, these are sent back excluding the receiver's information. This procedure is shown graphically in Fig.~\ref{fig:bp_messages}.

\subsubsection{Numerical results}
\begin{figure}
    \includegraphics[width=\linewidth]{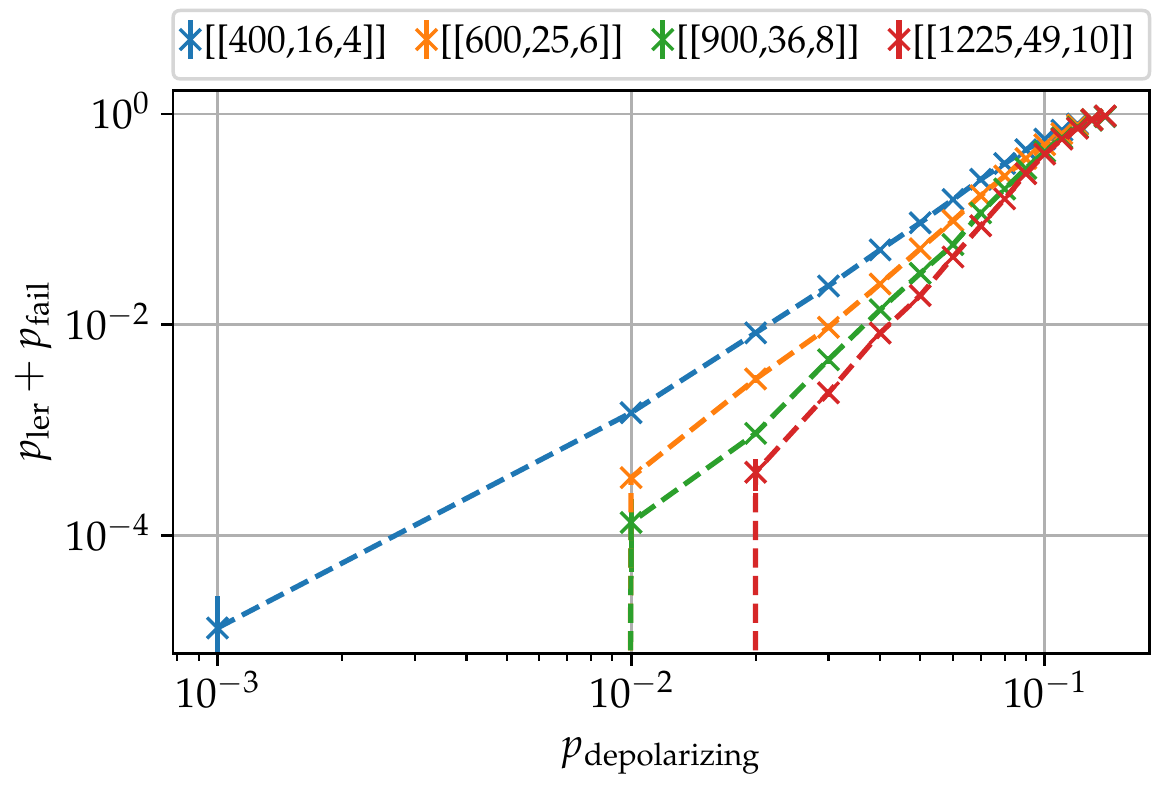}
    \includegraphics[width=\linewidth]{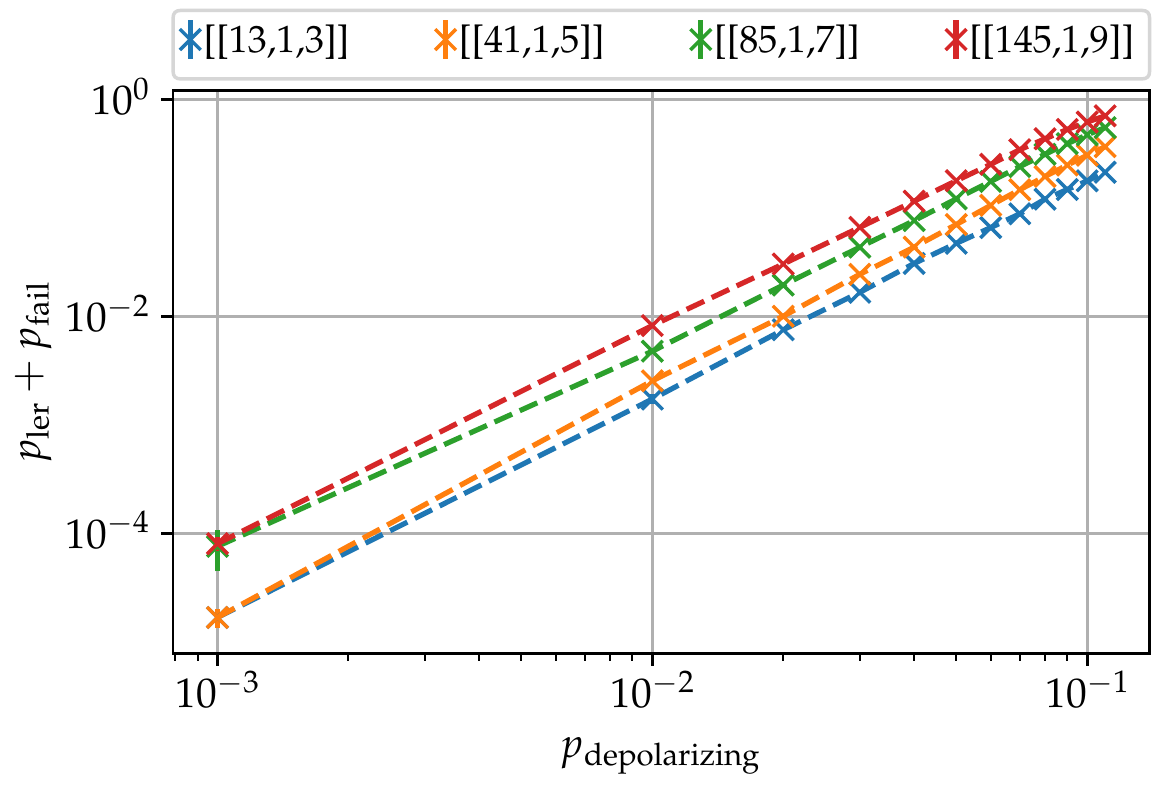}
    \caption{GBP with the Bethe approximation (standard BP), for HGP codes based on random classical codes (top) and topological surface codes (bottom). While the random codes show the emergence of a (pseudo-) threshold, the surface codes show decreasing decoding performance for increasing distance. \label{fig:bp_bethe}}
\end{figure}

Using this choice of regions, we obtain the decoding performance for hypergraph product codes based on random matrices and for topological surface codes shown in Fig.~\ref{fig:bp_bethe}. 
The random codes are \((7,4)\)-qLDPC and the surface codes are \((4,4)\)-qLDPC.
We see that the random codes show a good performance in the sense that increasing the distance of the code increasingly suppresses failures. The surface codes however show the opposite behavior. 
In both cases, the primary cause for a decoding failure is not a logical error, but a failure to return an error compatible with the syndrome or a failure of convergence.

\begin{figure}
    \centering
    \includegraphics[width=\linewidth]{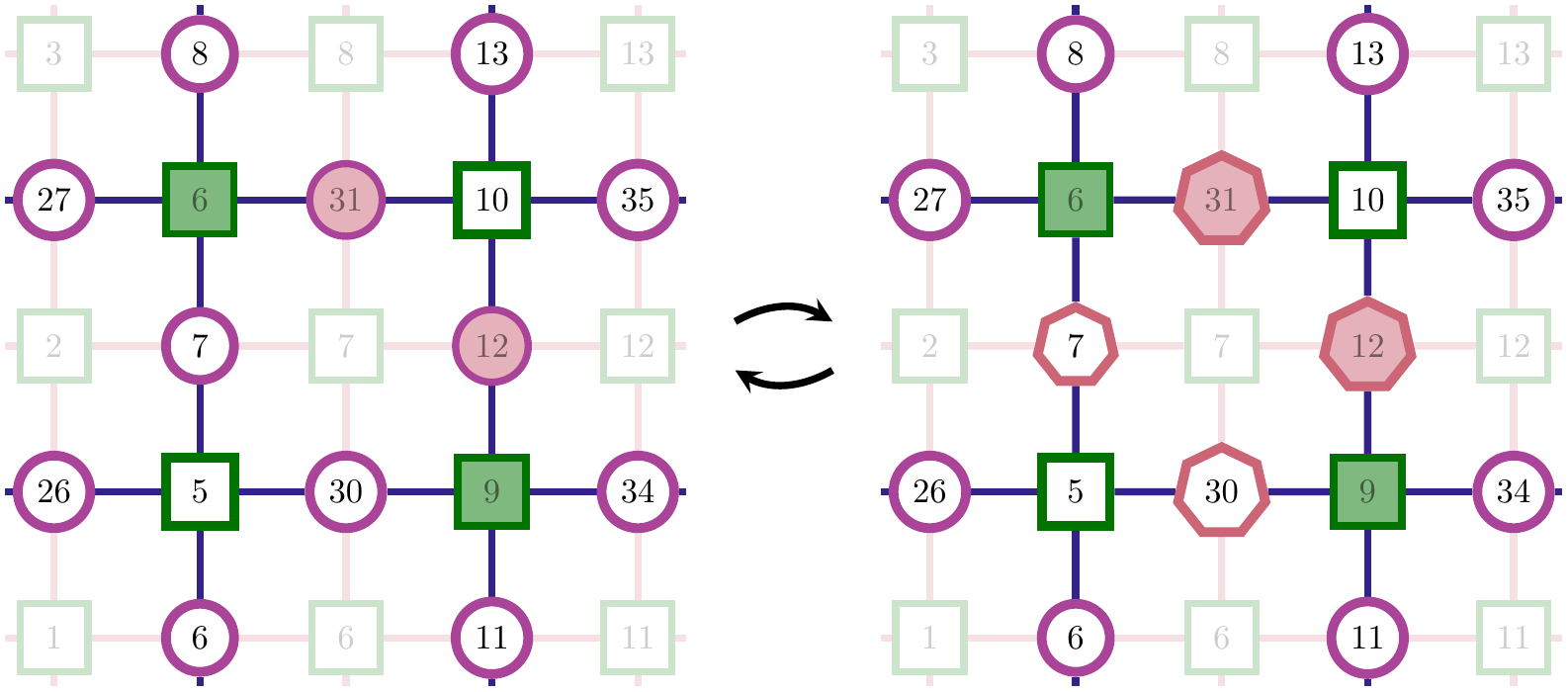}
    \caption{Split belief on a patch of the surface code, decoded with BP. Tanner graph representation with qubits as purple circles, parity-checks as green squares. \(X\)-Paulis are represented by blue and \(Z\)-Paulis by red edges. The initial error is indicated by filled qubits, the violated checks by filled squares. The error guess of the decoder by the heptagons. The \(Z\)-error on qubits \(Z_{12} Z_{31}\) violates parity checks \(\{6,9\}\). The (degenerate) error pattern \(Z_{7} Z_{30}\) is symmetric to the original one. The standard BP decoder returns the union of all those qubits \(Z_{7} Z_{30} Z_{12} Z_{31}\) with empty syndrome leading to a decoding failure.}
    \label{fig:split_belief}
\end{figure}

\subsubsection{Success and obstacles using BP}
The disadvantages and problems involved in classical BP decoding are extensively covered in the literature. These mainly concern harmful patterns in the Tanner graphs of the code due to cycles or  \emph{trapping sets}~\cite{price2017survey}. 
Their existence in classical codes translates to quantum codes when using constructions based on the classical codes, like the hypergraph product construction.
This also means that classical methods like message scheduling can be used to alleviate such problems~\cite{raveendran2021trapping}.
The nature of quantum codes themselves introduces new obstacles to BP decoding. 
Certain syndromes allow for different error configurations, that can be translated to each other by symmetry transformations in the corresponding Tanner graph.

In a qubit-wise decoding fashion, the BP decoder assigns the same probability to each qubit involved in such configurations. With this \emph{split belief}, the decoder thresholds all those qubits to the same error guess and therefore fails to converge. Some scheduling methods can break these symmetries but there is no general method to avoid them. Using irregular base codes and graphs of odd degree distribution also reduces the amount of symmetry, helping the decoder. An exemplary split belief is shown and explained in Fig.~\ref{fig:split_belief}.

Because surface codes are highly symmetrical, lots of such split beliefs occur during decoding, which explains their bad performance.
HGP codes with random base codes however show a good decoding performance because their local topology is inherited from the random classical codes that do not exhibit split beliefs.

There are two main strategies for improving the performance of BP. The first makes changes to the algorithm itself, \emph{Memory Belief Propagation} for example shows good decoding performance at the cost of a slightly higher complexity~\cite{kuo2021exploiting}. 
A version of GBP was used to improve the decoding performance in quantum bicycle codes~\cite{raveendran2019syndrome}.

Other methods use the \emph{soft} output of the BP algorithm (\textit{i.e.} the marginal probabilities) and use them as input for post-processing methods. Notably \emph{Ordered Statistics Decoding} allows to apply the combined BP+OSD decoder across a wide range of qLDPC codes, again at the cost of a higher complexity~\cite{panteleev2019degenerate,roffe2020decoding}.

\section{GBP for the Surface Code} \label{sec:GBP_surface}
\begin{figure*}
    \centering
    \includegraphics[width=0.8\linewidth]{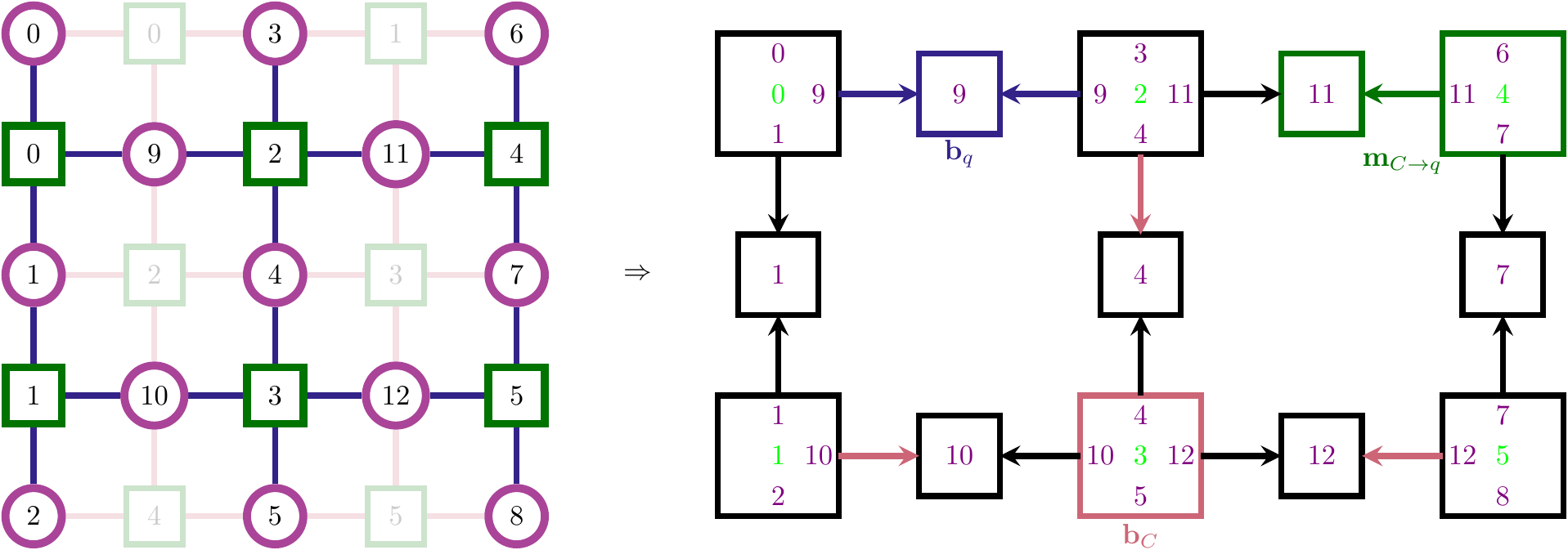}
    \caption{Tanner graph (left) to region graph (right) for the distance \(3\)-surface code. Indicated in the region graph are the contributions to large region beliefs (red), small region beliefs (blue) and messages (green).}
    \label{fig:rg_surface}
\end{figure*}

We use the same region graph as used for the Bethe approximation in Sec.~\ref{sss:bethe_approximation}, \textit{i.e.} with large and small regions. 
The reason for that choice is that for surface codes, this construction method is equivalent to using the \emph{cluster variational method}, a standard method for constructing valid region graphs~\cite{yedidia2005constructing}. This choice of regions ensures that the smallest split beliefs, i.e. those related to two syndrome excitations, are resolved. The transition from the Tanner graph to the region graph is shown for a binary implementation in Fig. \ref{fig:rg_surface}.

\begin{figure}
    \centering
    \includegraphics[width=\linewidth]{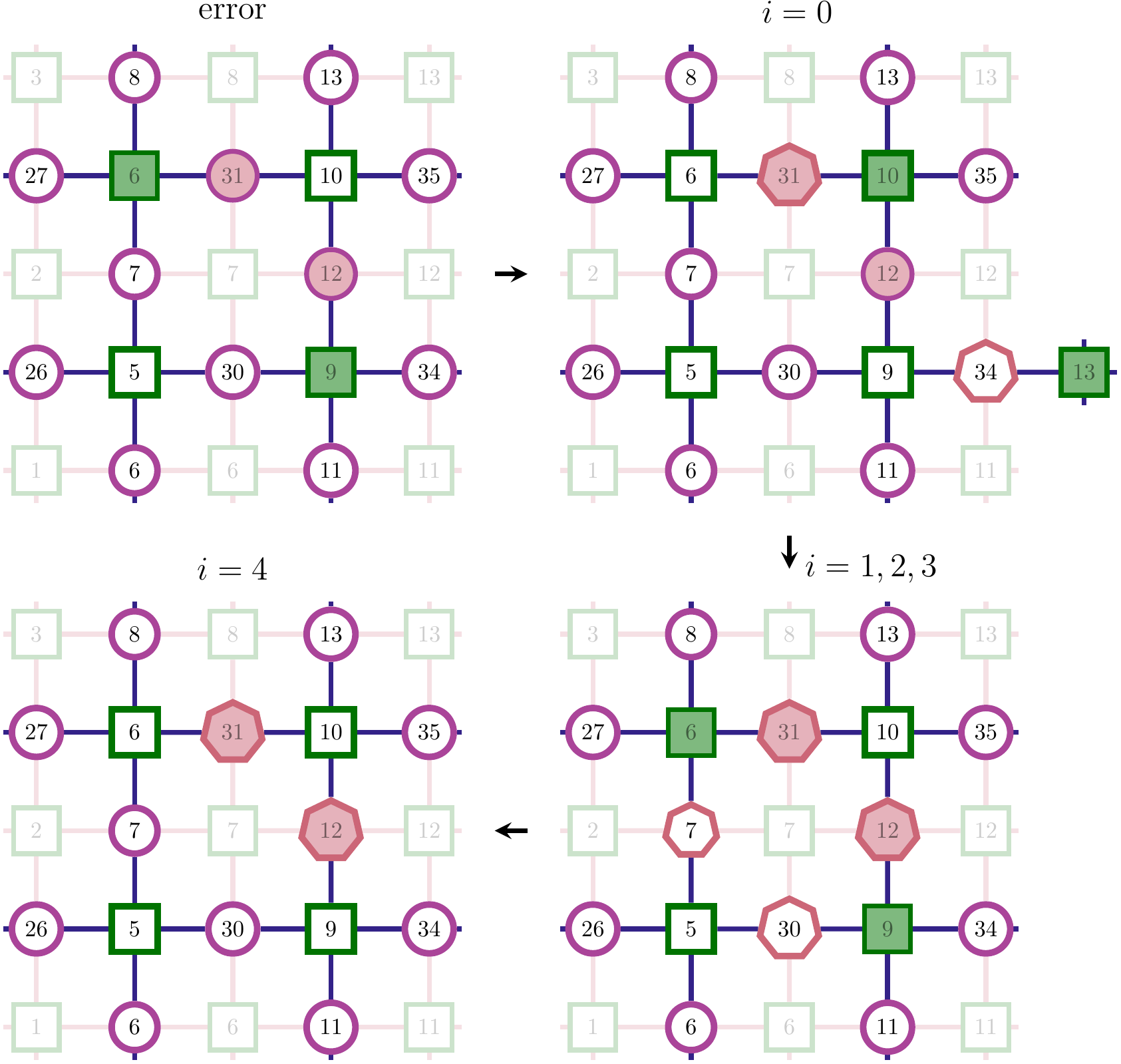}
    \caption{Split belief on a patch of the surface code, decoded with GBP. Tanner graph representation with qubits as purple circles, parity-checks as green squares. \(X\)-Paulis are represented by blue and \(Z\)-Paulis by red edges. The initial error is indicated by filled qubits, the violated checks by filled squares. The error guess of the decoder by the heptagons. The \(Z\)-error on qubits \(Z_{12} Z_{31}\) violates parity checks \(\{6,9\}\). The (degenerate) error pattern \(Z_{7} Z_{30}\) is symmetric to the original one in the sense that they have equal weight. After \(i = 4\) iterations, the GBP decoder returns the correct error, successfully decoding the split belief.}
    \label{fig:split_belief_gbp}
\end{figure}

We implement the hard decision based on Eq.~\ref{eqn:gbp_hard_decision}. We show how this helps with the prototypical split belief from Fig.~\ref{fig:split_belief} in Fig.~\ref{fig:split_belief_gbp}.
Using this hard decision procedure, we might still get a split belief within a single large region. However in our simulations, we observe that this is not the case, \textit{i.e.} the \(\argmax\) of the region is unique towards the end of the iterations.

\subsection{Split and Repeat}

We observe that while decoding based on the region beliefs using Eq.~\ref{eqn:gbp_hard_decision_2} improves upon standard BP, there still exist error guesses incompatible with the overall syndrome. However, applying the proposed correction usually reduces the syndrome weight.
This is similar to one decoding iteration in the small-set flip algorithm (SSF), proposed by Leverrier, Tillich and Zémor \cite{leverrier2015quantumexpander}. In the SSF-algorithm, a configuration of qubits in the support of a stabilizer is flipped, if it decreases the syndrome weight. This works well for quantum codes with a sufficiently expanding Tanner graph. In surface codes however there are constant weight error patterns, that lead to failure of the SSF-algorithm. Our hard decision heuristic after the GBP inference that reduces the residual syndrome weight suggests that this issue can be overcome.

\begin{figure}
    \includegraphics[width=\linewidth]{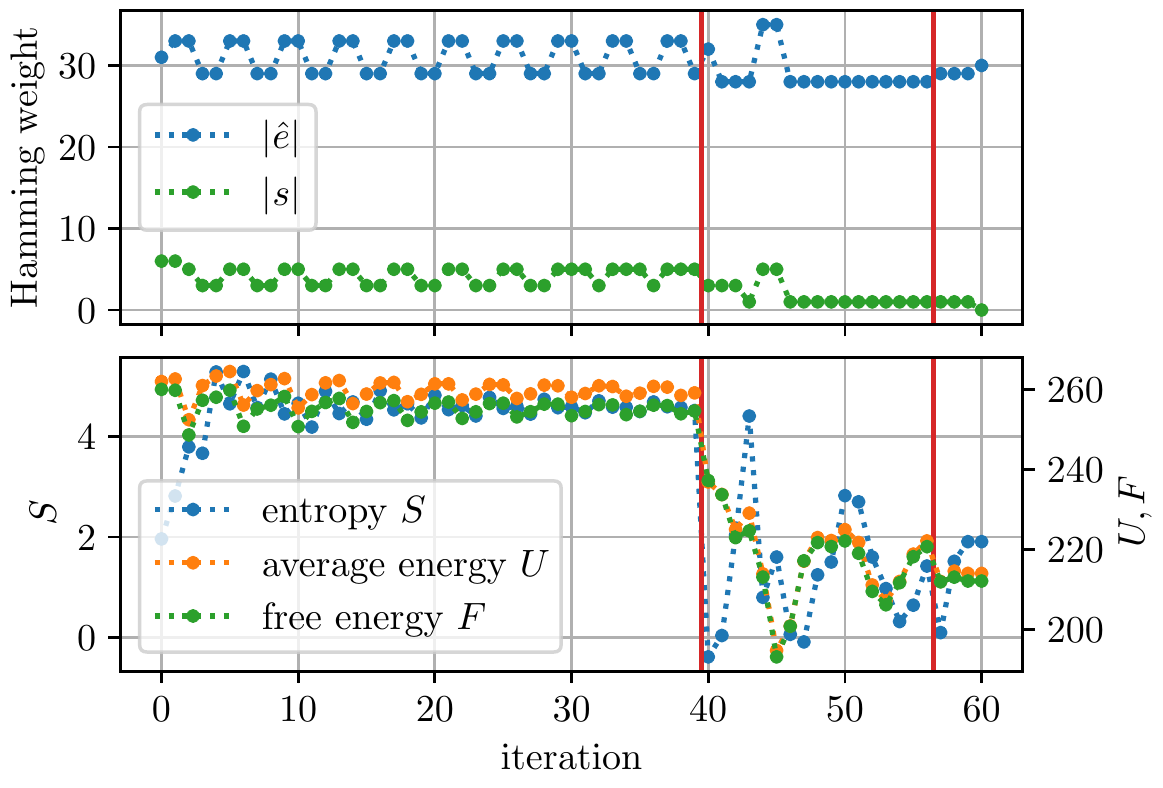}
    \caption{Weight of total error guess and residual syndrome (top) and thermodynamic quantities (bottom) during decoding of an error on a distance \(9\) surface code. Vertical lines represent the re-initialization after convergence or maximum number of iterations reached. We see that we can escape the oscillatory behavior by re-initializing. \label{fig:gbp_freeenergy}}
\end{figure}

We use this insight to formulate a \emph{split and repeat} procedure:  
After a run of GBP, we save the current error guess and reinitialize the decoding procedure with the syndrome of lower weight and a rescaled error probability. We repeat this until there is an empty overall syndrome. 
The overall error guess then is the sum of all intermediate errors. This algorithm is shown in Alg.~\ref{alg:gbp_rs_surface}. An exemplary run for a particularly harmful error pattern on a distance \(9\) surface code is shown in Fig.~\ref{fig:gbp_freeenergy}. We also see that the free energy is reduced during the course of decoding correspondingly.

\begin{algorithm}
        \SetAlgoLined
        \DontPrintSemicolon
        \BlankLine
        \KwIn{Parity-Check matrix \(\vb{H}\), syndrome \(s\),\\ a-priori probability \(p_{\text{init}}\), maximum number of iterations and repetitions \(n_{\text{mi}}, n_{\text{mr}}\)}
        \KwOut{Error guess \(\hat{e}\)}
        \(\mc{RG} = \) region graph constructed from \(\vb{H}\) with Bethe approximation \;
        \(i = 0\) \;
        \(\hat{e}_{\text{total}} = 0\) \;
        \(s_i = s\) \;
    
        \While{\(i  < n_{\text{mr}}\)}
        {    
            \(\tilde{p}_{\text{init}} = \abs{p_{\text{init}}-\mathrm{wt}(\hat{e}_{\text{total}}) / n_{\text{qubits}}}\)\;
            \( \hat{e}_{i} = \mathrm{GBP}(s_{i}, \mc{RG}, \tilde{p}_{\text{init}}, n_{\text{mi}}) \) \;
            \( \hat{e}_{\text{total}} = \hat{e}_{\text{total}} + \hat{e}_{i}\) \;
            \eIf{\( \sigma(\hat{e}_{\text{total}}) \ifed s_{\mathrm{GBP}}  = s \)}
            {
                \KwRet{\(\hat{e} = \hat{e}_{\text{total}}\)}
            }
            {
                \(s_{i+1} = s_i + s_{\mathrm{GBP}}\) \;
                \(i =  i + 1 \)
            }
        }
        \KwRet{Fail}

        \caption{\(\mathrm{GBP}\) split repeat decoding for surface codes.}
        \label{alg:gbp_rs_surface}
\end{algorithm}
    
\subsection{Dependence on Initial Probabilities}
An additional parameter of the decoding procedure is the initial probability. As mentioned by Hagiwara et al., a linear decoder with a fixed initialization can correct a certain error set, independent of the error probability ~\cite{hagiwara2012fixed}. Kuo and Lai remark that choosing a fixed initialization can prevent fluctuations and increase decoding stability~\cite{kuo2021exploiting}. 
In our simulations, we observe that a fixed initialization error probability can lead to decoding failure. We therefore consider two further adaptions.

On the one hand, when re-initializing after a split procedure, we rescale the channel error probability by the weight of the current error guess, see Alg.~\ref{alg:gbp_rs_surface}.

On the other hand, when decoding still fails, we reinitialize the whole decoder with different initial probabilities.
Assuming a good decoding performance when \(p_{\text{init}}\) is close to the channel error probability, we sample the new initial probability from a Gaussian with width \(0.1\) around the channel error probability. 

\subsection{Numerical Results}
\begin{figure}[h]
\centering
    \includegraphics[width=\linewidth]{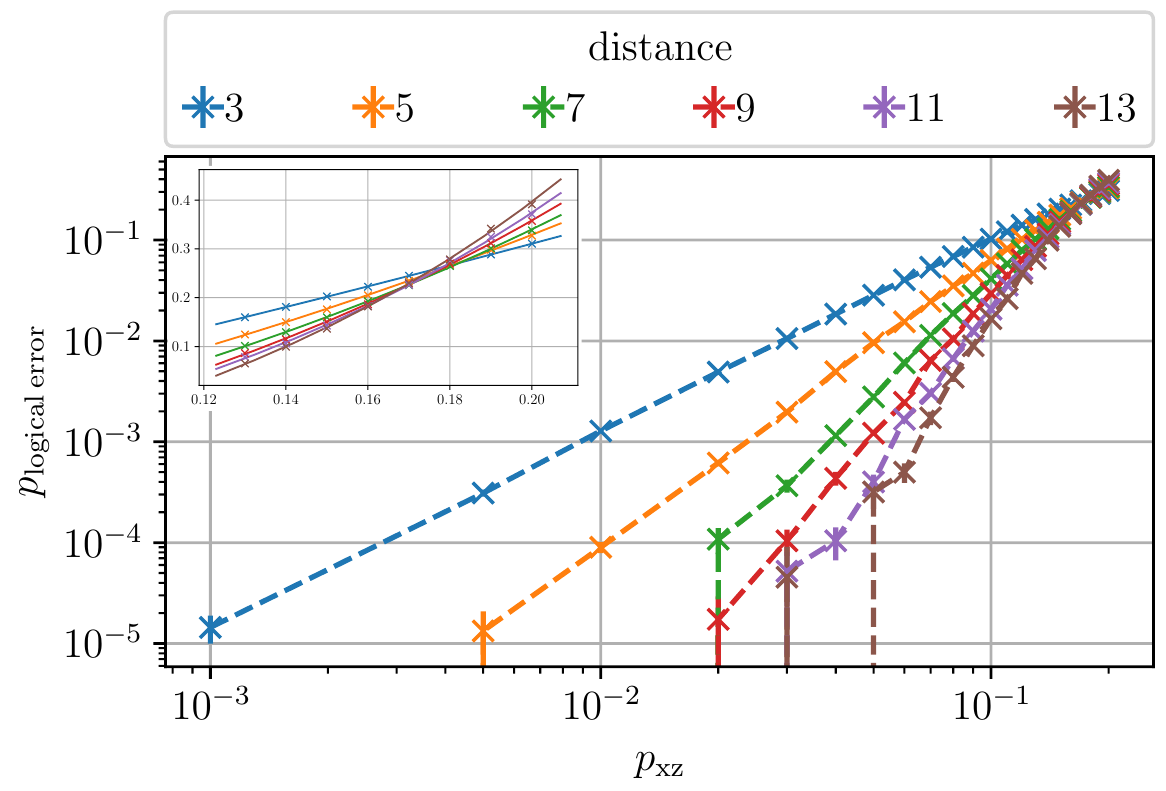}
    \includegraphics[width=\linewidth]{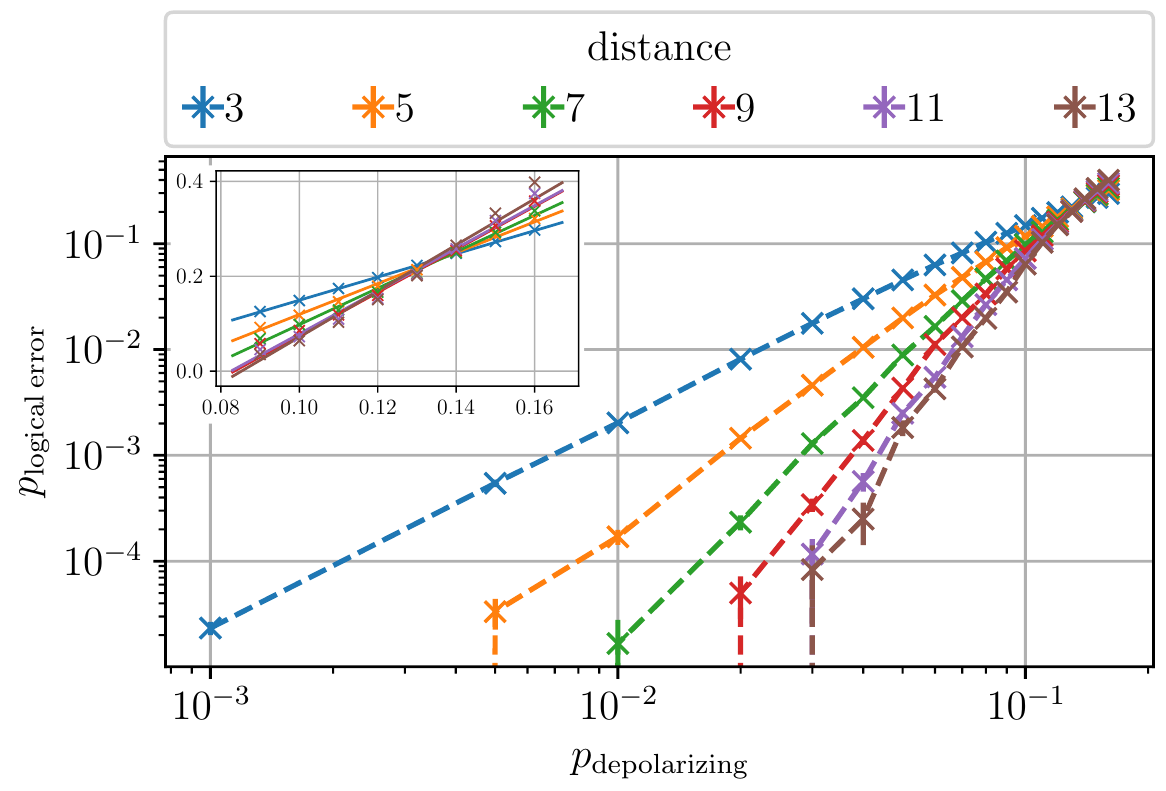}
    \caption{GBP for the surface code for independent \(XZ\)-noise (top) and depolarizing noise (bottom) in binary implementation. The logical error rate decreases with increasing distance. We estimate thresholds of \(p_{\mathrm{th}}^{\mathrm{XY}} \approx 17 \%\) and  \(p_{\mathrm{th}}^{\mathrm{depol.}} \approx 13.3 \%\).\label{fig:gbp_surface}}
\end{figure}

\begin{figure}[h]
\centering
    \includegraphics[width=\linewidth]{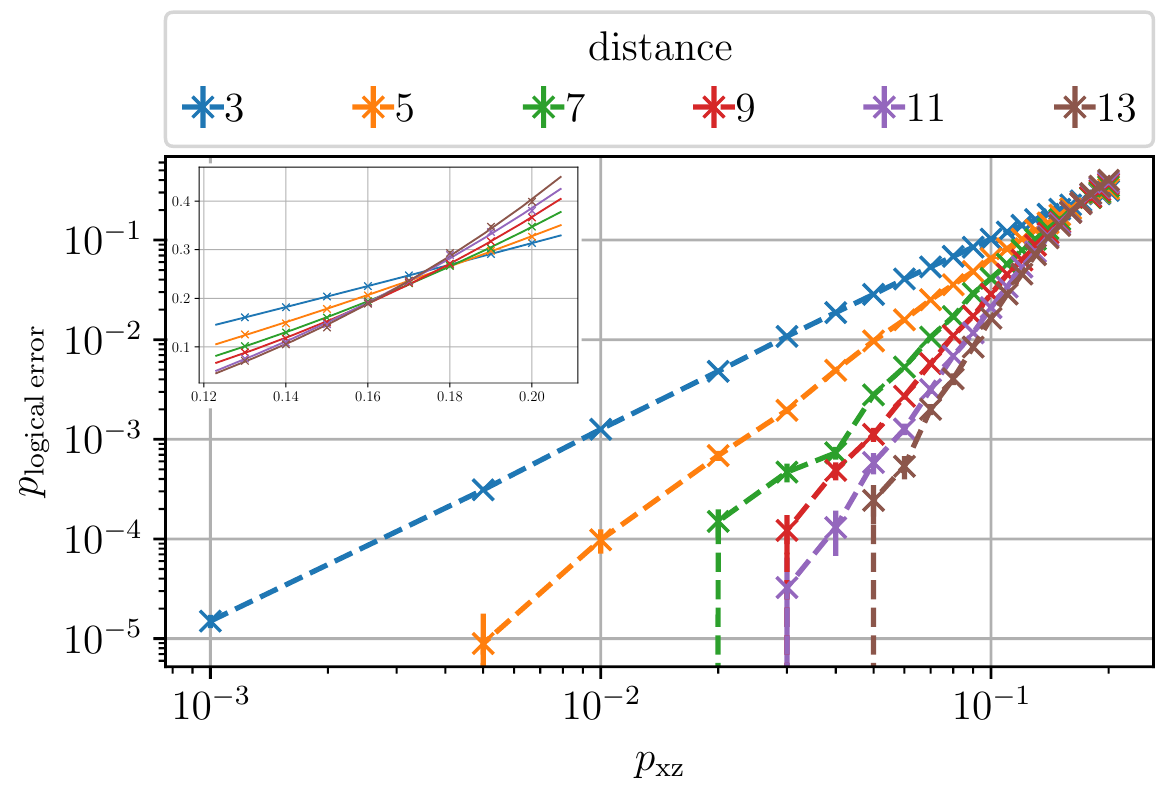}
    \includegraphics[width=\linewidth]{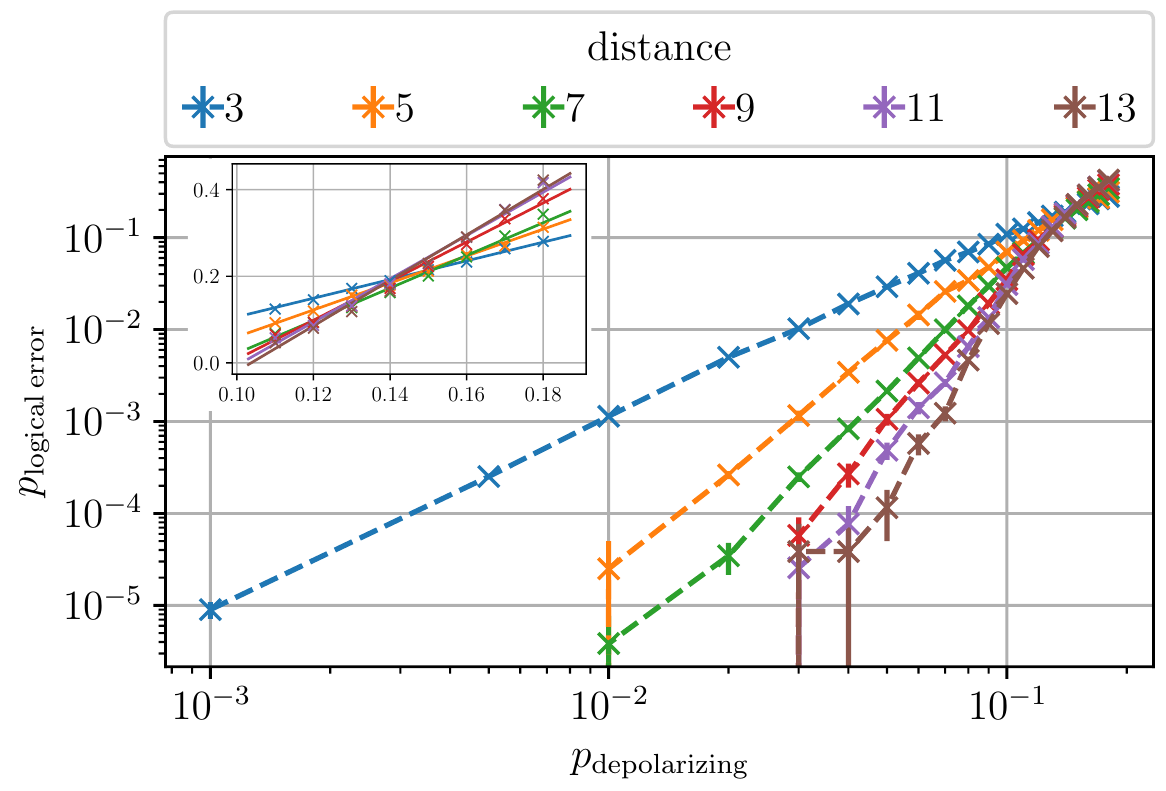}
    \caption{GBP for the surface code for independent \(XZ\)-noise (top) and depolarizing noise (bottom) in quaternary implementation. The logical error rate decreases with increasing distance. We estimate thresholds of \(p_{\mathrm{th}}^{\mathrm{XY}} \approx 17 \%\) and  \(p_{\mathrm{th}}^{\mathrm{depol.}} \approx 14 \%\). \label{fig:gbp4_surface}}
\end{figure}

Remarkably, in our simulations we never observe a decoding failure and the decoder always returns an error guess that puts the corrupted word back to the codespace. 
The results are shown for independent \(XZ\)- noise and depolarizing noise in Fig.~\ref{fig:gbp_surface} and ~\ref{fig:gbp4_surface} for the binary and quaternary implementation respectively. 
They generally show a decreasing logical error probability with increasing distance and a crossing indicating a threshold. The results for the threshold are summarized in Tab.~\ref{tab:gbp_surf_results}.

\begin{table}%
\centering
    \caption{Thresholds from error sampling on the \(XZ\)-channel and the depolarizing channel for binary and quaternary implementation. The thresholds fall short of the optimal thresholds obtained by statistical mechanic methods but are similar to the ones from BP-OSD decoding. The optimal threshold for the \(XZ\)-channel is obtained from the single-Pauli threshold \(p_{\mathrm{th}}^X \approx 10.9 \%\) as \(p_{\mathrm{th}}^{XZ} = 2 p_{\mathrm{th}}^X - (p_{\mathrm{th}}^X)^2 \).  \label{tab:gbp_surf_results}}

        \begin{tabularx}{\linewidth}{ccccc}
        \toprule
            Ch.                       & \(q\) & \(p_{\mathrm{th}}\) & BP-OSD & optimal   \\
            \midrule
            \multirow{2}{*}{\(XZ\)}       & \(2\) & \(17 \%\)           & 17.6\%~\cite{roffe2020decoding}  & \multirow{2}{*}{ \(20.6 \%\)~\cite{dkpl2001topological}} \\
                                          & \(4\) & \(17\% \)  &  &                                                          \\
            \midrule
            \multirow{2}{*}{depol.} & \(2\) & \(13.3 \%\)         &  & \multirow{2}{*}{ \(18.9 \%\)~\cite{bombin2012strong}}    \\
                                          & \(4\) & \(14\% \) &  &                                                          \\
            \bottomrule
        \end{tabularx}

\end{table}

\paragraph*{Complexity}
In a naive implementation, directly adapting Eqs.~\ref{eqn:region_belief_general} and~\ref{eqn:gbp_message_update}, the GBP algorithm requires
\begin{itemize}
    \item \(q^{d_c}\) multiplications per check region \(c\) \(\to \leq n_c q^{d_c}\),
    \item \(q^{d_q}\) multiplications per qubit region \(q\) \(\to \leq n_q q^{d_q}\),
    \item \(q^{d_c - 1}\) summations per marginalization of check region \(c\) \(\to \leq n_c (q^{d_c - 1})\),
    \item \(q\) multiplications and divisions per message calculation \(\to \leq 2 d_q n_q q\),
\end{itemize}
amounting to an overall asymptotic complexity of \(\mc{O}(n_{\text{repetitions}}n_{\text{iterations}} q^{d_c} n_c)\).  By implementing parallel the calculation of beliefs and messages, the explicit dependence on the code size can be omitted,  \(\mc{O}(n_{\text{repetitions}}n_{\text{iterations}})\). The amount of repetitions and iterations needed still depends on the code size and heuristically scale as shown in Tab.~\ref{tab:gbp_surf_scaling}.
They amount to an overall scaling of \(\mc{O}(n_c^{2})\) (GF(2)) and \(\mc{O}(e^{\sqrt{n_c}} n_c^{2})\) (GF(4)). %

\begin{table}%
\centering
    \caption{Scaling of iterations and repetitions of the decoder with code distance  \label{tab:gbp_surf_scaling}, heuristically obtained from simulations.}
        \begin{tabularx}{\linewidth}{ccc}
        \toprule
                                               & \(\mathrm{GF}(2)\)             & \(\mathrm{GF}(4)\) \\
            \midrule
            Iterations                         & \(\mc{O}(d^2) = \mc{O}(n_c)\)  & \(\mc{O}(d^2) = \mc{O}(n_c)\)       \\
            Split rep.                 & \(\mc{O}(d^2)= \mc{O}(n_c)\)   & \(\mc{O}(d^2)= \mc{O}(n_c)\)       \\
            \(p_{\mathrm{init}} \) rep. & \(\mc{O}(d^0)= \mc{O}(1)\) & \(\mc{O}(\exp(d))\)       \\
        \bottomrule
        \end{tabularx}
\end{table}

\begin{figure}
    \includegraphics[width=0.48\linewidth]{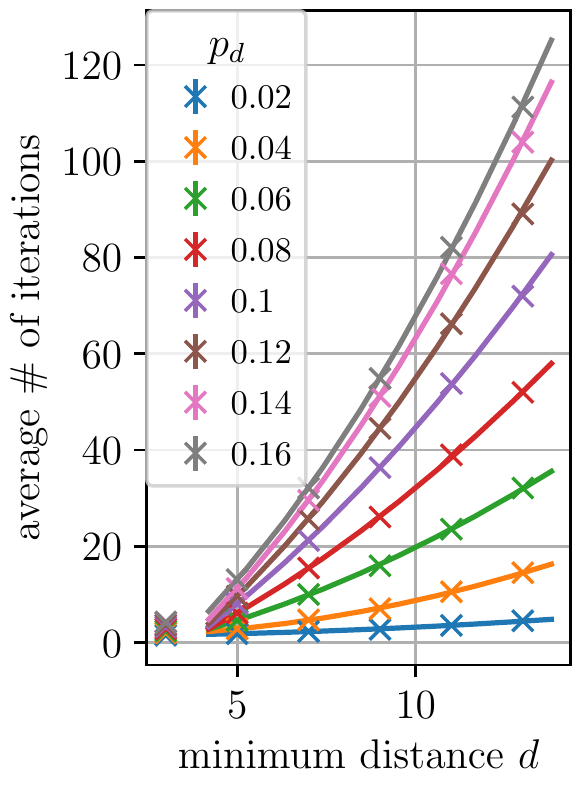}
    \includegraphics[width=0.48\linewidth]{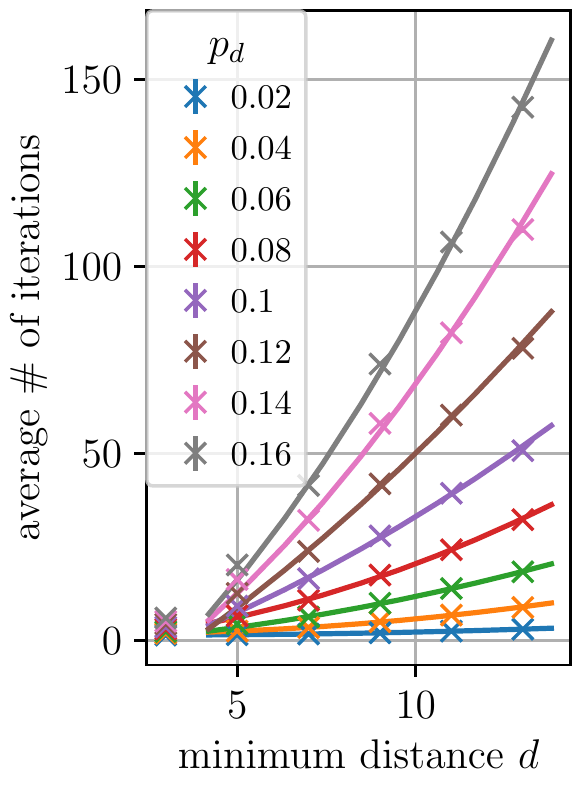}
    \caption{Average number of inner iterations scales quadratic with the code distance, shown is a quadratic fit for distances \(d \geq 5\). (Left) Binary implementation in the depolarizing channel. (Right) Quaternary implementation in the depolarizing channel. The quaternary implementation needs on average more iterations. \label{fig:gbp_scaling_iterations}}
\end{figure}

\begin{figure}
    \includegraphics[width=0.48\linewidth]{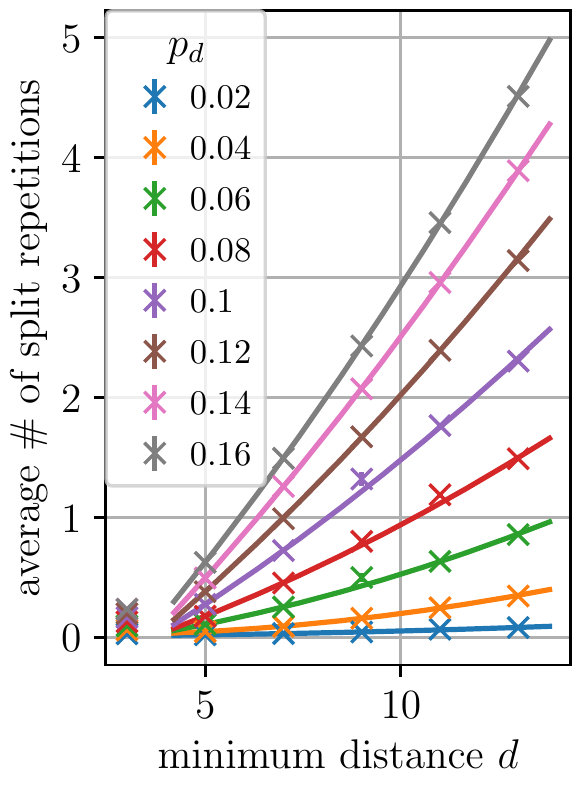}
    \includegraphics[width=0.48\linewidth]{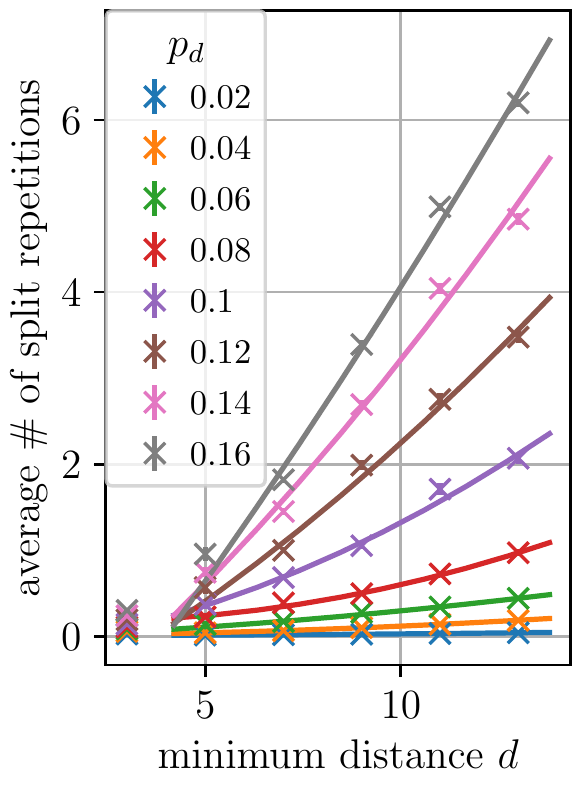}
    \caption{Average number of split repetitions scales quadratic with the code distance, shown is a quadratic fit for distances \(d \geq 5\). (Left) Binary implementation in the depolarizing channel. (Right) Quaternary implementation in the depolarizing channel. The quaternary implementation needs on average more repetitions. \label{fig:gbp_scaling_reps}}
\end{figure}

\section{Summary and Outlook}
We developed a decoder based on Generalized Belief Propagation using a specific hard decision method and an outer re-initialization loop. With these adaptations, the decoder is able to decode the surface code. The logical error probability decreases with growing distance below a certain qubit error probability indicating the emergence of a threshold of about \(14\%\) for depolarizing noise and \(17\%\) for independent bit- and phase-flip noise. As is typical for practical decoders, these values fall short of theoretical upper bounds but offer a lower decoding complexity. When comparing to known decoding algorithms, our decoder shows similar threshold values compared to BP-OSD. The main ingredient to the OSD-post processing is matrix inversion, which scales with the third power in the number of rows, \textit{i.e.} \(\mc{O}(n_c^3)\)~\cite{roffe2020decoding}. 
Minimum weight perfect matching achieves a higher threshold but scales in general as \(\mc{O}(n_q^3)\)~\cite{kolmogorov2009blossom}.
The almost linear time Union find decoder also achieves a slightly higher threshold~\cite{delfosse2021almost}.

Future work can include a lower complexity implementation that is based on log-likelihood-ratios, which is frequently used in standard BP algorithms to reduce complexity. This would allow the decoder to be tested for more general quantum LDPC codes, where finding a fast and general decoder is ongoing research.

Additionally, all simulations were performed with code capacity noise, \textit{i.e.} the data qubits experience noise through the quantum channel and the syndrome readout is assumed perfect. A next step on the road towards fault tolerance is to extend the decoding scheme to more realistic noise models, for example including faulty syndrome measurements. There are recent results suggesting that belief propagation is also suitable for syndrome noise when using repeated measurements and even in a single-shot decoding scheme~\cite{grospellier2020combining,quintavalle2021single}.

\section*{Simulation Methods}
The decoder is implemented in a \(\mathrm{GF}(q)\) formalism with both \(q = 2\) and \(q = 4\) in C++ making use of libraries libDai~\cite{mooij2010libDAI}, the lemon Graph library~\cite{dezso2011lemon}, xtensor~\cite{mabille2016xtensor}, NTL~\cite{shoup2021libntl} and nlohmann JSON headers~\cite{nlohmann2022json}. The code can be found on github~\cite{old2022gbp}.

\section*{Acknowledgements}
We would like to thank B.M. Terhal and J. Knörzer for comments on the manuscript and M. Müller and D.P. DiVincenzo for facilitating this project. We acknowledge support from the EU Quantum Technology Flagship grant AQTION under grant agreement
number 820495 and by the BMBF project MUNIQC-ATOMS. This research
is also part of the Munich Quantum Valley (K-8),
which is supported by the Bavarian state government with funds from the Hightech Agenda Bayern Plus. MR was supported by ERC grant EQEC No. 682726 during the initial part of this work. Simulations were performed with computing resources granted by RWTH Aachen University under project \emph{thes1045}. 
\bibliographystyle{quantum}
\bibliography{gbp_paper_bib}

\appendix

\section{Variational Methods in Statistical Mechanics} \label{app:Variational}
The paradigmatic example of a variational method in quantum mechanics is the Ritz method~\footnote{colloquially often known as Rayleigh-Ritz}. Here, the setting is that we are given a Hamiltonian and the task is to find its ground-state. Since this task is in general computationally hard already for the simplest non-trivial practically relevant Hamiltonians, it is fruitful and instructive to develop systematic methods to construct trial wavefunctions that approximate the ground state wavefunction while retaining computational feasibility. The fundamental insight here is that the energy expectation value of the problem Hamiltonian evaluated on \emph{any} state is lower bounded by the expectation value of the true ground state,
\begin{align}
\bra{\psi_\mathrm{trial}} \hat H \ket{\psi_\mathrm{trial}} \geq \bra{E_0} \hat H \ket{E_0},
\end{align}
which essentially only relies on the fact that we can expand $\ket{\psi}$ in the Hamiltonian eigenbasis, upon which the statement follows immediately.\\

This paradigm of systematically constructing trial states can be extended to mixed states and finite temperature. The role of energy is taken over by the (Helmholtz) free energy $F=E-TS = -\frac{1}{\beta}\log Z$, where $S$ is the entropy and $Z=\tr\exp(-\beta \hat H)$ the partition function. With respect to the free energy, any trial state fulfils the Bogoliubov inequality
\begin{align}
\mathrm{tr} \hat F \rho_\mathrm{trial} \geq \mathrm{tr} \hat F \rho_\mathrm{c}.
\end{align}
Note that this contains the Rayleigh-Ritz inequality in the zero temperature limit. It can be proved by exploiting the Gibbs inequality
\begin{align}
\tr \left[A \log A - A\log B \right] \geq 0,
\end{align}
which holds for any $A,B \geq 0$ provided $\tr A = \tr B$. Plugging in the canonical ensemble (or Gibbs state) $\rho_\mathrm{c} = \exp(-\beta \hat H)/Z$ for $B$ leads to
\begin{align}
- S(\rho_\mathrm{trial}) + \beta \tr \hat H \rho_\mathrm{trial} + \log Z \geq 0, \label{eqn:var}
\end{align}
which can be rearranged into the Bogoliubov inequality, stating that the free energy of any trial density matrix $\rho_\mathrm{trial}$ is lower bounded by the free energy of the canonical ensemble. This permits the extension of the variational principle to mixed states, where we now try to approximate the Gibbs state by a trial density matrix. In the context of the present work, we are dealing ``only'' with probability distributions, so let us point out the rather trivial fact that the above inequality in particular also holds when $\hat H$ is diagonal and the density operators are simply multivariate probability distributions.

\subsection{Minimization of the Free Energy}
For a variational ansatz, we introduce the trial probability distribution, the \emph{belief} \(b(\vb{x})\). Equivalent to Eq.~\ref{eqn:var}, it holds that the free energy of such a trial probability distribution is lower bounded by the free energy of the real distribution, \textit{i.e.}
\begin{align}
    F(b) \defi U(b) - S(b) &\geq F_H \\
    \sum_{\vb{x} \in P} b(\vb{x}) E(\vb{x}) + \sum_{\vb{x} \in P} b(\vb{x}) \ln b(\vb{x}) &\geq -\ln{Z}.
\end{align}
Here, \(U(b)\) is the \emph{(variational) average energy} and \(S(b)\) the \emph{(variational) entropy} of the trial state.
Plugging in the definition of the energy gives
\begin{align}
    F(b) = F_H + \sum_{\vb{x} \in P} b(\vb{x}) \ln \frac{b(\vb{x})}{p(\vb{x})} = F_H + D(b||p)
\end{align}
with \(F(b) \geq F_H\) and equality iff \(b(\vb{x}) = p(\vb{x})\). \(D(b || p)\) is the Kullback-Leibler divergence that gives a measure of how close the trial distribution \(b\) is to the "true" distribution \(p\). 
Because \(D(b || p) \geq 0\) with equality iff \(b(\vb{x}) = p(\vb{x})\), a minimization procedure of \(D(b||p)\) with respect to \(b(\vb{x})\) can exactly compute \(F_H\) and recover \(p(\vb{x})\). 

Due to the intractability of a brute force approach, the trial functions are generally restricted or approximated by some factorized form. %In the following, we review a general approach to construction approximations, that include the standard BP (\emph{Bethe}-) approximation.

\section{Representation of Stablizer Codes \label{app:representation}}
\paragraph{Binary representation \(q = 2\)}.
The binary representation maps Pauli words of length \(n\) to binary vectors of length \(2n\). We can represent any Pauli word (up to a phase) by \(E = X^{\vb{e}_x} Z^{\vb{e}_z}\), where \(\vb{e}_x,\vb{e}_z \in \mathrm{GF}(2)^{n}\). The binary representation then are the concatenations \(\vb{H} = (\vb{H}_X, \vb{H}_Z)\) and \(\vb{e} = (\vb{e}_x,\vb{e}_z)\) and it holds 
\begin{align}
        E_i = 
        \begin{cases}
                X \qif \vb{e}_i = 1 &\qand \vb{e}_{i+n} = 0, \\
                Y \qif \vb{e}_i = 1 &\qand \vb{e}_{i+n} = 1, \\
                Z \qif \vb{e}_i = 0 &\qand \vb{e}_{i+n} = 1.
        \end{cases} 
\end{align}
The addition in \(\mathrm{GF}(2)\) corresponds to addition modulo \(2\), \(1 + 1 = 0\). The symplectic product of to vectors \(\vb{e} = (\vb{e}_x,\vb{e}_z), \vb{e}' = (\vb{e}'_x,\vb{e}'_z) \) is defined via
\begin{align*}
        \vb{e} \star \vb{e}' \defi \vb{e} &\vb{P} \vb{e}' =  \vb{e}_x \cdot \vb{e}'_z + \vb{e}_z \cdot \vb{e}'_x\\
        \qwith &\vb{P} \defi 
        \left(\begin{matrix}
                0 & \iden_n \\
                \iden_n & 0 
        \end{matrix} \right) \qand \cdot %\text{ the scalar product},
\end{align*}
the scalar product, such that 
\begin{align}
        \vb{H} \star \vb{e}  =  \vb{H} &\vb{P} \vb{e}. \label{eqn:sym_gf2}
\end{align}
\paragraph{Quaternary representation \(q = 4\)}.
In \(\mathrm{GF}(4)\), addition and multiplication can be defined via its addition and multiplication tables
\begin{align}
        \renewcommand{\arraystretch}{0.4}
        \begin{minipage}{0.5\linewidth}
                \[\begin{array}{c|cccc}
                        + & 0 & 1 & \omega & \bar{\omega} \\
                        \hline
                        0 & 0 & 1 & \omega & \bar{\omega} \\
                        1 & 1 & 0 & \bar{\omega} & \omega \\
                        \omega & \omega & \bar{\omega} & 0 & 1 \\
                        \bar{\omega} & \bar{\omega} & \omega   & 1 & 0
                \end{array}\]
        \end{minipage}
        \begin{minipage}{0.5\linewidth}
                \[\begin{array}{c|cccc}
                        \times & 0 & 1 & \omega & \bar{\omega} \\
                        \hline
                        0 & 0 & 0 & 0 & 0 \\
                        1 & 0 & 1 & \omega & \bar{\omega} \\
                        \omega & 0 & \omega & \bar{\omega} & 1 \\
                        \bar{\omega} & 0 & \bar{\omega}  & 1 & \omega
                \end{array}\]
        \end{minipage}.
\end{align}
The group operation is naturally represented by addition, if the Paulis are mapped via
\begin{align}
        I \to 0 \qc X \to 1 \qc Y \to \overline{\omega} \qc Z \to \omega.
\end{align}
In order to define the symplectic product, two more definitions are needed, the conjugation and trace in \(\mathrm{GF}(4)\),
\begin{itemize}
        \item Conjugation: \(\mathrm{GF}(4) \to \mathrm{GF}(4) : \alpha \to \overline{\alpha} = \alpha \times \alpha\),
        \item Trace: \(\mathrm{GF}(4) \to \mathrm{GF}(4) : \alpha \to \Tr{\alpha} = \alpha + \overline{\alpha} \).
\end{itemize}
Then
\begin{align}
        \vb{e} \star \vb{e}' \defi \Tr{ \vb{e} \cdot \vb{e}'} = \Tr{ \sum_q \vb{e}_q \times \overline{\vb{e}'_q}}
\end{align}
such that 
\begin{align}
       \vb{s} = \vb{H} \star \vb{e}  =  \left( \Tr{ \sum_q \vb{H}_{cq} \times \overline{\vb{e}_q}} \right)_{c=0}^{n-k-1}. \label{eqn:sym_gf4}
\end{align}

\section{Hypergraph Product Construction \label{app:hgp}}
The Tanner graph of the hypergraph product (quantum) code \(\mc{T}_{\mc{Q}}\) is based on the Cartesian product of the classical Tanner graphs \(\mc{T}_{\mc{C}_1}\) and \(\mc{T}_{\mc{C}_2}\).
        The Cartesian product of two graphs \(\mc{T}_1 = (N_1 = V_1 \cup C_1,E_1)\) and \(\mc{T}_2 = (N_2 = V_2 \cup C_2,E_2)\) is the bipartite graph \(\mc{T}_{1 \times 2} \ifed \mc{T}_1 \times \mc{T}_2 = (N_{1 \times 2},E_{1 \times 2})\) with
        \begin{align}
                N_{1 \times 2} &= \{n_1 n_2 | n \in N_1, n_2 \in N_2\} \\
                E_{1 \times 2} &= \{ (n_1 n_2, n'_1 n'_2) | (n_1 = n'_1 \wedge (n_2,n'_2) \in E_2)  \nonumber\\
                &\phantom{ = \{ (n_1 n_2, n'_1} \; \vee (n_2 = n'_2 \wedge (n_1,n'_1) \in E_1) \}.
        \end{align}
        In words: the vertex set of the resulting graph is the Cartesian product of the vertex sets of the graph factors. There is an edge between vertices in the resulting graph if any of their partial vertices shared an edge in their graph factor.
The graph constructed from the Cartesian product of two bipartite graphs is again bipartite. In order to derive a code, the vertex set of the new graph is partitioned into \(N_{1 \times 2} = Q \cup (C_X \cup C_Z)\) with
\begin{itemize}
    \item qubits: \(Q \defi \{n_1 n_2 | (n_1 \in V_1 \wedge n_2 \in V_2) \vee (n_1 \in C_1 \wedge n_2 \in C_2)\}\)
    \item \(X\)-type stabilizers: \(C_Z \defi \{n_1 n_2 | (n_1 \in C_1 \wedge n_2 \in V_2)\}\)
    \item \(Z\)-type stabilizers: \(C_X \defi \{n_1 n_2 | (n_1 \in V_1 \wedge n_2 \in C_2)\}\)
\end{itemize}
Chosen like that, the graph corresponds to a Tanner graph of a quantum CSS code. The commutation condition is fulfilled since whenever \(n_i \in V_i\) is adjacent to \(n'_i \in C_i\) in \(\mc{T}_{\mc{C}_i}\), there are exactly two vertices (qubits) in \(\mc{Q}\) which are adjacent to the constructed \(X\)-type stabilizer \(n'_i n_j\) and \(Z\)-type stabilizer \(n_i n'_j\): \(n'_i n'_j\) and \(n_i n_j\). Twofold anti-commutation then gives commutation.

\section{Belief Propagation Equations \label{app:bp}}
We denote by \(\Gamma(\bullet)\) the neighbors of node \(\bullet\) and by \(\sigma(\bullet)\) the parity of configuration \(\bullet\). For a detailed description of the steps, see text.
\begin{itemize}
    \item Qubit to check messages 
    \begin{align}
        \displaystyle m^{(i+1)}_{q \to c} \propto p_0(E_q) \prod_{c' \in \Gamma(q) \setminus c} m^{(i)}_{c' \to q}(E_q) \label{eqn:bpq2c}
    \end{align}
    \item Check to qubit messages
    \begin{align}
        m^{(i)}_{c \to q} \propto \sum_{\vb{E}_{\Gamma(c) \setminus q}} \delta[\sigma(\vb{E})_c = s_c] \prod_{q' \in \Gamma(c) \setminus q} m^{(i)}_{q' \to c}(E_{q'}) \label{eqn:bpc2q}
    \end{align}
    \item Belief
    \begin{align}
        b^{(i)}_{q}(E_q) \propto p_0(E_q) \prod_{c \in \Gamma(q)} m^{(i)}_{c \to q}(E_q)
    \end{align}
    \item hard decision / error guess
    \begin{align}
        \hat{E}^{(i)}_q = \argmax_{E_q} b^{(i)}_q(E_q)
    \end{align}
\end{itemize}

\end{document}